\documentclass[journal,onecolumn,]{IEEEtran}
\usepackage{cite}

\ifCLASSINFOpdf
   \usepackage[pdftex]{graphicx}
   \DeclareGraphicsExtensions{.pdf,.jpeg,.png}
\else
\fi
\usepackage{amsmath}
\usepackage{array}

\usepackage{caption}
\usepackage{subcaption}
\usepackage{algorithm}
\usepackage{algpseudocode}

\usepackage{newfloat}
\DeclareFloatingEnvironment[
  fileext=frm,
  placement={!ht},
  name=Listing
]{listing}
\DeclareCaptionSubType{listing}

\usepackage{color}
\usepackage{soul}

\usepackage{epstopdf}

\usepackage{listings}
\definecolor{comments}{rgb}{0.8,0,0}
\definecolor{numbers}{rgb}{0.0, 0.5, 0.0}
\definecolor{keywords}{rgb}{0.0,0.0,0.5}
\definecolor{strings}{rgb}{0.58,0,0.82}
\definecolor{background}{gray}{0.85}
\definecolor{addkeywords}{rgb}{0.0, 0.0, 1.0}
\DeclareCaptionStyle{listing}[justification=justified]{}

\usepackage{xcolor}
\usepackage{xparse}
\NewDocumentCommand{\framecolorbox}{oommm}
 {
  \IfValueTF{#1}
   {%
    \IfValueTF{#2}
     {\fcolorbox{#3}{#4}{\makebox[#1][#2]{#5}}}
     {\fcolorbox{#3}{#4}{\makebox[#1]{#5}}}%
   }
   {\fcolorbox{#3}{#4}{#5}}%
 }

\usepackage{url}

\begin{document}
%
\title{OpenCL/CUDA algorithms for parallel decoding \\ of any irregular LDPC code using GPU}
%
%
%

\author{Jan Broulim, Alexander Ayriyan, Vjaceslav Georgiev, Hovik Grigorian
\thanks{Jan Broulim with the University of West Bohemia, Univerzitni 22, 306 14 Pilsen and the Institute of Experimental and Applied Physics, Czech Technical University in Prague, Horska 3a/22, 128 00, Praha 2, Czech Republic. (e-mail: broulim@kae.zcu.cz).}
\thanks{Alexander Ayriyan is with the Laboratory of Information Technologies, Joint Institute for Nuclear Research, Joliot-Curie 6, 141980 Dubna, Russia (e-mail: ayriyan@jinr.ru).}
\thanks{Vjaceslav Georgiev is with the University of West Bohemia, Univerzitni 22, 306 14 Pilsen, Czech Republic (e-mail: georg@kae.zcu.cz).}
\thanks{Hovik Grigorian is with the Laboratory of Information Technologies, Joint Institute for Nuclear Research, Joliot-Curie 6, 141980 Dubna, Russia (e-mail: hovik.grigorian@gmail.com).}
}
\maketitle

\begin{abstract}
The development of multicore architectures supporting parallel data processing has led to a paradigm shift, which affects communication systems significantly. This article provides a scalable parallel approach of an iterative LDPC decoder, presented in a tutorial-based style. It is suitable for decoding any irregular LDPC code without the limitation of the maximum node degree, and it includes a parallel calculation of the syndrome. This is the main difference from algorithms presented so far. The proposed approach can be implemented in applications supporting massive parallel computing, such as GPU or FPGA devices. The implementation of the LDPC decoder with the use the OpenCL and CUDA frameworks is discussed and the performance evaluation is given at the end of this contribution. 
\end{abstract}


%
\IEEEpeerreviewmaketitle

\section{Introduction}
%
%
%
%
Since Shannon's work, the topic of error detection and error correction codes, related to channel coding, has seen significant growth \cite{djordevic_2016}. The first serious discussion of error correction codes emerged in Hamming's work in 1950 \cite{hamming_1950}, where Hamming provided a method for the correction of single and the detection of double bit errors with minimum redundancy being added to the transmitted data. Since the second half of the 20th century, error correction codes have attracted much attention in research work and have been utilized in many applications, including deep space photography transmission \cite{spacecraft_2008}, television broadcasting services \cite{dvb_2009}, Ethernet \cite{ethernet_2007}, wireless communication networks, and other signal processing applications. 

This paper provides a parallel approach of an iterative Low Density Parity Check (LDPC) \cite{gallager_1963, bonello_2011, wilberg_1996} decoder, presented in a tutorial style. The presented parallel approach can be implemented in platforms allowing massive parallel computing, such as Graphics Processing Units (GPUs),  Field Programmable Gate Arrays (FPGAs), and computer data storages. The proposed approach is not limited for certain families of LDPC codes, but it supports decoding of any irregular LDPC code, and the maximum node degree is not limited. Benchmarks  of the LDPC decoder implemented using Open Computing Language (OpenCL) \cite{opencl_spec} and Compute Unified Device Architecture (CUDA) \cite{cuda_spec} frameworks are discussed and the performance comparison is given at the end of this contribution. 

This contribution can be easily used as a tutorial for implementing an irregular LDPC decoder as well as a general parallel approach for additional optimizations in order to make further accelerations. The parallel decoding approach is suitable for fast decoders implemented in GPUs. It is also highly applicable for accelerating Bit Error Rate simmulations used in designing new LDPC codes.

Inspiring by various comparisons between the OpenCL and CUDA applications from different fields of research, e. g. \cite{opencl1,opencl2,opencl3}, we developed parallel algorithms for LDPC decoding using OpenCL and CUDA. Several contributions published so far deal with a general comparison of OpenCL and CUDA \cite{performance_2011} and with fitting the LDPC decoder on GPU platform \cite{gpu_1,gpu_2,gpu_3,gpu_4,gpu_5,gpu_7,gpu_8,gpu_9,gpu_10,gpu_11}. However, the decoders are mostly limited for applications with some families of LDPC codes or bounded with the maximum node degree in the associated Tanner graph \cite{tanner_1981}. The proposed parallel approach is suitable for decoding any irregular LDPC code without the bound in terms of the maximum node degree.

\hfill \today

\section{LDPC}
\subsection{Introduction}
LDPC codes \cite{gallager_1963} represent the coding technique with the best known error correcting capabilities. LDPC codes surpassed other codes \cite{spielman_1998}, including turbo codes \cite{turbo_1993} and Reed Solomon codes \cite{rs_1960}, at the correcting performance, and they are becoming increasingly difficult to ignore in novel communication signal processing systems. Although the number of applications with LDPC codes has grown significantly with the increasing speed of computing resources, decoding is still a computionally intensive task, which limits the deployability of non-approximated decoding algorithms for medium and long block length codes. However, the decoding can be accelerated significantly with the use of parallel multicore computing architectures. Our work related to LDPC codes include \cite{our_1,our_2,our_3,our_4,our_5}.


\begin{figure}[h!]
    \centering
        \includegraphics[width=0.5\textwidth]{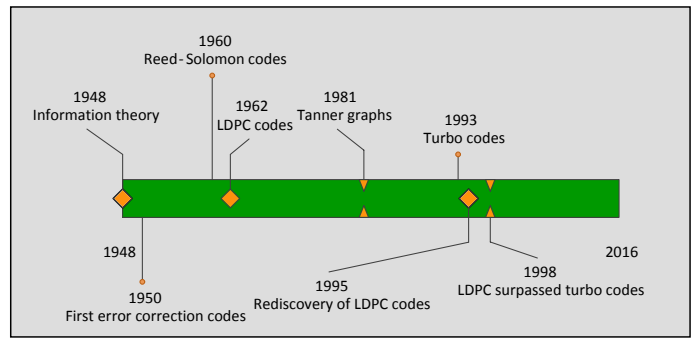}
        \caption{Historical milestones in error correction coding theory.}
        \label{fig:history}
\end{figure}
\begin{figure}[h!]
    \centering
        \label{fig:com_process}
        \includegraphics[width=0.5\textwidth]{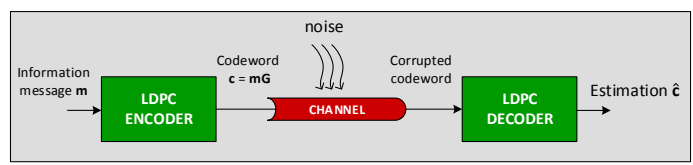}
        \caption{Communication process - transfering information through the noisy channel.}
        \label{fig:com_process}
\end{figure}

\begin{figure}[h!]
    \centering
        \includegraphics[width=0.5\textwidth]{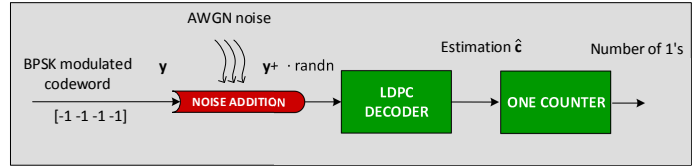}
        \caption{Utilization of the noisy channel for the Bit Error Rate simulation.}
        \label{fig:ber}
\end{figure}

\subsection{Basic defitions}

In this section, we provide basic mathematical definitions related to channel coding and their associations to LDPC codes and the presented parallel decoder.

Let $\mathcal{C}=(n,k)$ be a linear block code, where the number of code bits is denoted as $n$ and the number of information bits is denoted as $k$. The information vector of $k$ bits is denoted as $\mathbf{m}$ and the ($kn$) generator matrix is denoted as $\mathbf{G}$. The codeword $\mathbf{c}$ is given by $\mathbf{c}=\mathbf{mG}$, which is encoding. The parity-check matrix associated with the code $\mathcal{C}$ is denoted as  $\mathbf{H}$. Any vector $\mathbf{v}$ is a codeword if and only if $\mathbf{vH}^T=\mathbf{0}$. The product of the multiplication $\mathbf{vH}^T$ is called the syndrome. If the parity-check matrix $\mathbf{H}$ of code $\mathcal{C}$ is sparse, the code  $\mathcal{C}$ is said to be the Low-Density Parity-Check (LDPC) code. 

The Tanner graph is a bipartite graph of sets of variable nodes and check nodes defined by the parity-check matrix $\mathbf{H}$. If the element $H_{i,j} = 1$ ($i$ corresponds to the row, while $j$ corresponds to the column of the matrix  $\mathbf{H}$), an edge occurs between the check node $c_i$ and the variable node $v_j$. The Tanner graph is used for LDPC decoding, which is briefly described in the following section.

The vector of check nodes connected with $j$-th variable node is denoted as $\mathcal{M}_j$ be and the vector of variable nodes connected with the $i$-th check node is denoted as $\mathcal{N}_i$. Then
\begin{equation}
 \mathcal{M}_j = \{i\} \Leftrightarrow \mathbf{H}_{i,j} = 1
\end{equation}
\begin{equation}
 \mathcal{N}_i = \{j\} \Leftrightarrow \mathbf{H}_{i,j} = 1
\end{equation}

\subsection{Decoding}
Decoding is a method for correcting errors in a corrupted codeword and the device performing decoding is called the decoder. The output of the decoder is usually called the estimation ${\mathrm{\widehat{c}}}$, as illustrated in Fig.~\ref{fig:com_process}. Two main principles, listed below, can be considered for LDPC decoding. The principles are:

\begin{itemize} 
 \item Hard-decision, e. g. Bit-Flipping
 \item Soft-decision, working with probabilities during decoding process
\end{itemize}
Soft-decision decoding, including the Sum-Product (SP) algorithm~\cite{wilberg_1996} and its derivations, is supposed for the implementation of the LDPC decoder and related benchmarks in this article.

LDPC decoding is an iterative process of passing values as messages in the Tanner graph through its edges. An estimation of the codeword is calculated after finishing each iteration and if the estimation is a codeword of the LDPC code, decoding is stopped. If a codeword is not found after a certain number of iterations (typically 5-100), decoding is terminated as unsuccessful.

All messages passed in the Tanner graph represent probabilities, which are used for calculating the estimation after finishing every iteration. Because the algorithm convergence is affected significantly by the parameters of the Tanner graph (especially the number of short cycles), there is no reason for performing relatively high number of iterations. Therefore, the maximum number of iterations is limited. 

\begin{figure*}[t]  
    \centering
    \begin{subfigure}[b]{0.45\textwidth}
        \centering
        \includegraphics[width=\textwidth]{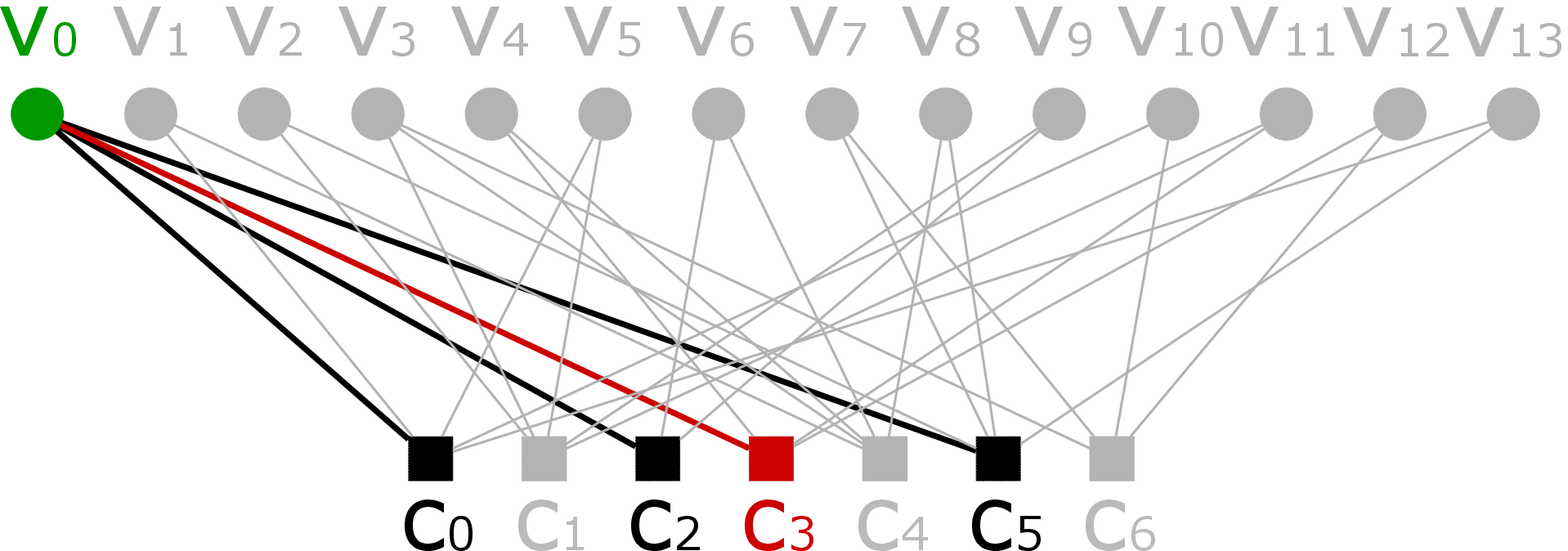}
        \caption{The first half of the iteration - from variable nodes to check nodes. Values used for the calculation of the message between $v_0$ and $c_3$ are highlighted.}
        \label{fig:iter_first_half}
    \end{subfigure}
    ~
    \begin{subfigure}[b]{0.45\textwidth}
        \centering
        \includegraphics[width=\textwidth]{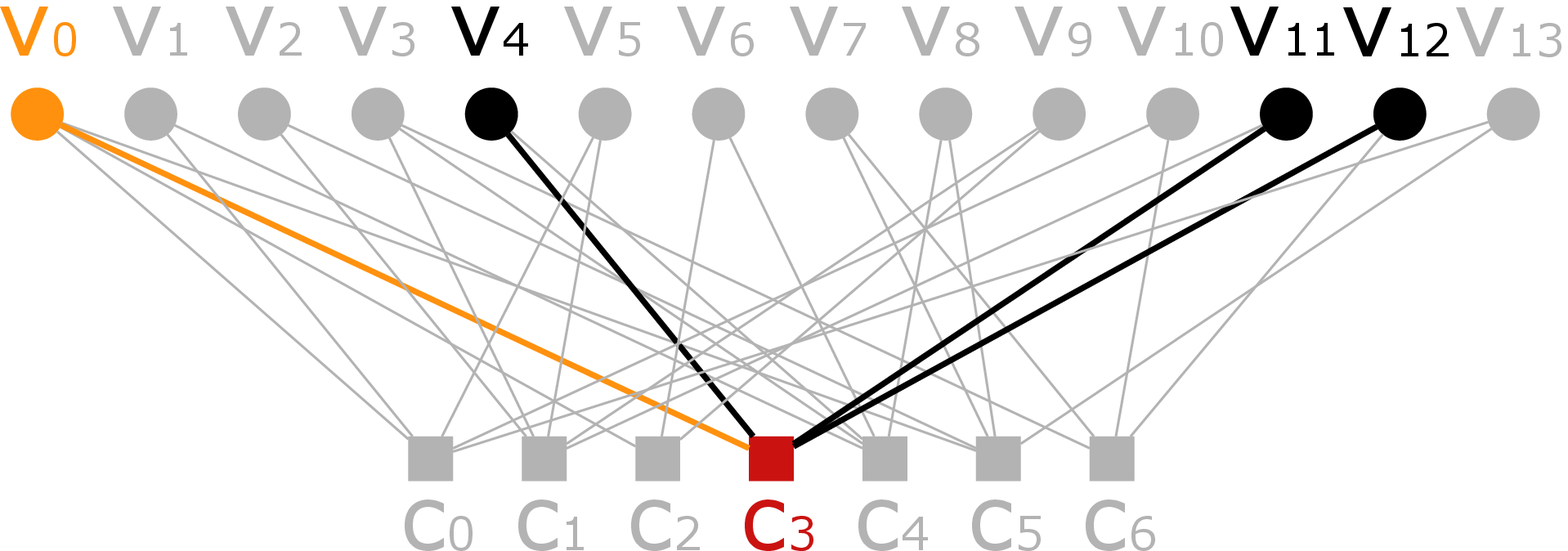}
        \caption{The second half of the iteration - from check nodes to variable nodes. Values used for the calculation of the message between $c_3$ and $v_0$ are highlighted.}
        \label{fig:iter_second_half}
    \end{subfigure}
    \caption{Tanner graph of the LDPC (14,7) code.}
    \label{fig:iter}
\end{figure*}

\begin{figure}[t!]
    \centering
        \includegraphics[width=6cm]{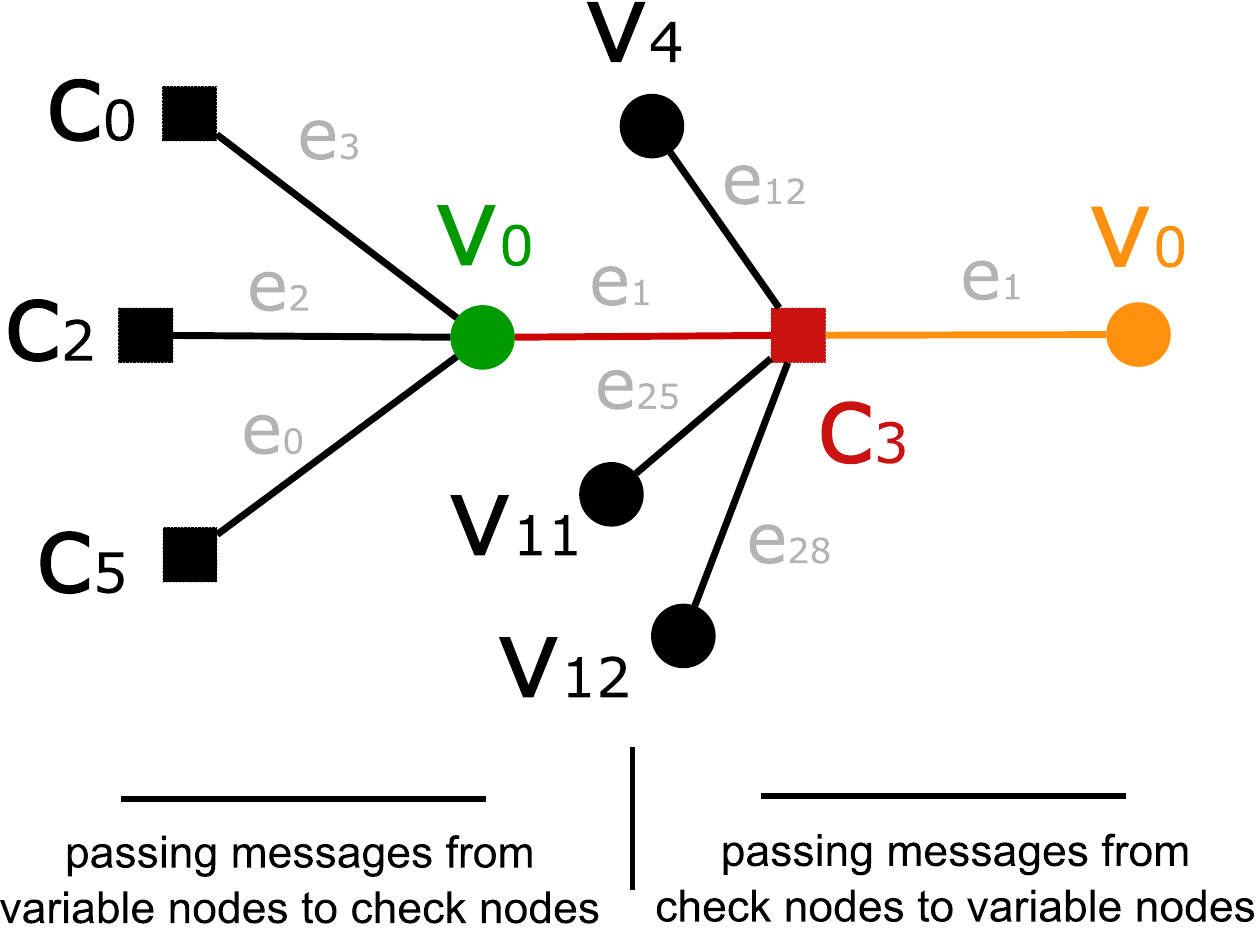}
        \caption{Data flow in the Tanner graph when passing values from $v_0$ to $c_3$ and back.}
        \label{fig:DataFlow}
\end{figure}

Messages outgoing from one set of nodes are calculated with the use of the incoming values from the opposite set of nodes. Edges are used as interfaces for passing messages between the set of variable nodes and the set of check nodes, while each message outgoing from a node is passed through an edge. Each message outgoing from a node in the Tanner graph depends on the incoming messages from the connected nodes excluding the value received from the node which is the destination node, as Algorithm 1 describes in more detail. The process is ilustrated in the following example. As can be seen in Fig.~\ref{fig:iter}, the variable node $v_0$ is connected with check nodes $c_0, c_2, c_3, c_5$. Considering the calculation of the value being passed from $v_0$ to $c_0$, the value depends on the incoming values from the nodes $c_2$, $c_3$ and $c_5$. In the second half of an iteration, the value being passed from $c_3$ to $v_0$ depends on the incoming values from $v_4, v_{11}, v_{12}$. The data flow is shown in Fig.~\ref{fig:DataFlow}. The passed values are used for calculating estimations after each iteration.

Soft-decision decoding, described in terms of the pseudocode, is listed in Algorithm~\ref{alg:decode} and in referenced Algorithm~\ref{alg:MessagePass}. Formulas used in the pseudocode represent the SP algorithm \cite{wilberg_1996} without any simplifications and modifications.

%
%

%
%

\begin{figure}[t]
    \centering
    \begin{subfigure}[b]{0.45\textwidth}
        \includegraphics[height=3.5cm]{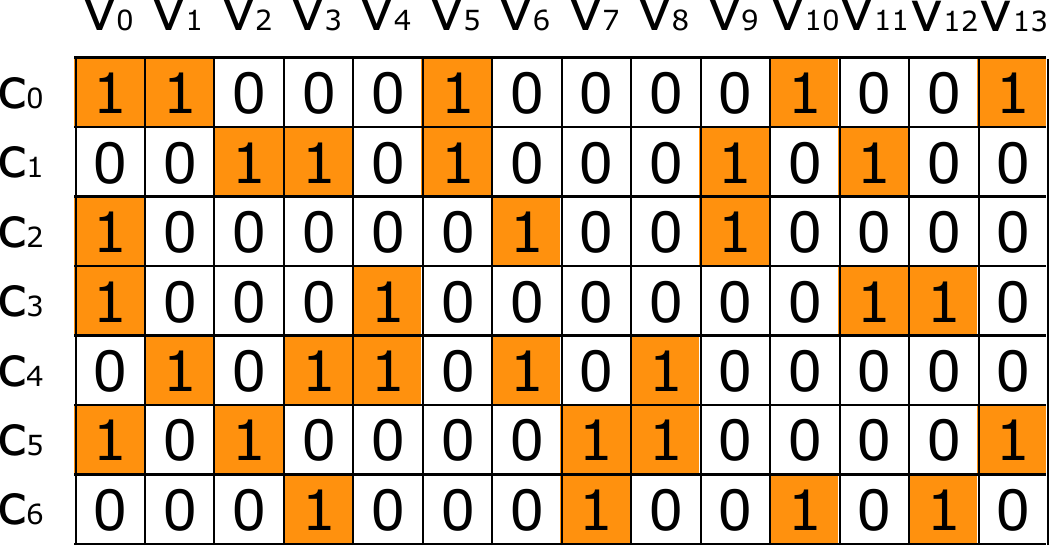}
        \caption{Parity-check matrix}
        \label{fig:ParCheckMatrix1}
       \vspace{2ex}
        ~\includegraphics[height=3.5cm]{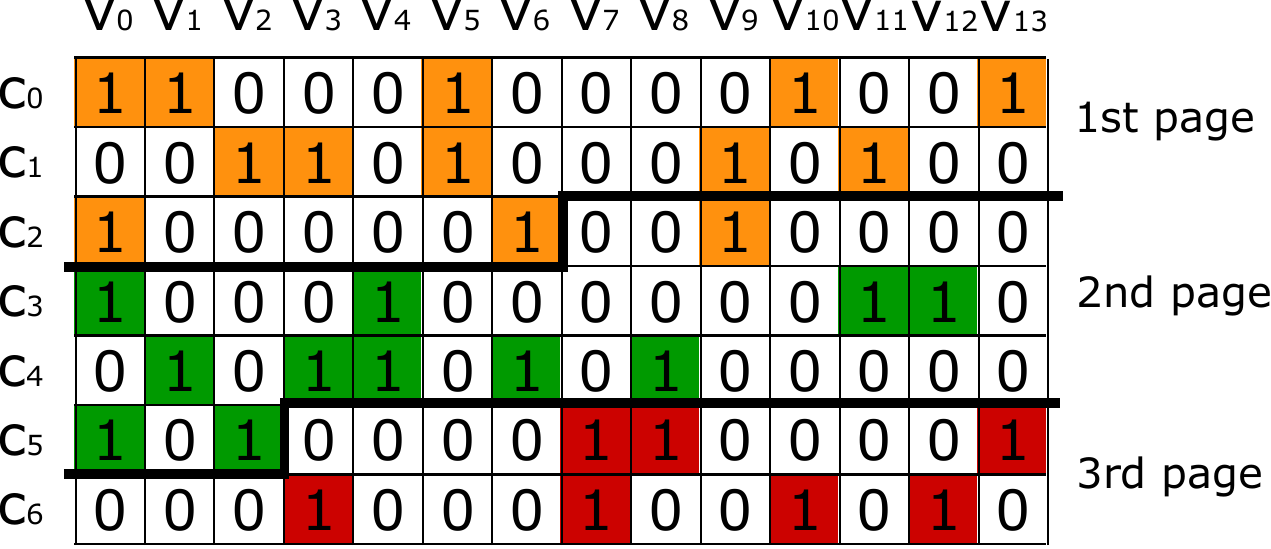}
        \caption{Parity-check matrix divided into pages}
        \label{fig:ParCheckMatrix2}
    \end{subfigure}
    \caption{Parity-check matrix and the principle of the parallelization}
    \label{fig:ParCheckMatrix}
\end{figure}

%
%

\begin{algorithm}[h!]
\renewcommand{\algorithmicrequire}{\textbf{\quad\,\, Input:}}
\renewcommand{\algorithmicensure}{\textbf{\quad\,\, Output:}}
\caption{Message passing}
\begin{algorithmic}[1]

\Procedure {Values to Check Nodes}{}
\newline
\Comment{First half on an iteration}
  \Require $\mathbf{p}$, $\mathbf{r}$
  \Ensure $\mathbf{q}$

  \ForAll {$j \in \mathbf{[0,|\mathcal{M}|)}$}
  \ForAll {$i \in \mathbf{[0,|\mathcal{N}|)}$}
      \State $q_{i,j}^0 =p_j^0 $
      \State $q_{i,j}^1 =p_j^1 $
      \ForAll {$i' \in {\mathcal{M}_j  \setminus  i}$}
	    \State $q_{i,j}^0 =q_{i,j}^0 r_{i',j}^0 $
	    \State $q_{i,j}^1 =q_{i,j}^1 r_{i',j}^1 $
      \EndFor
  \EndFor
  \EndFor

\EndProcedure

\Statex

\Procedure {Values to Variable Nodes}{}
\newline
\Comment{Second half on an iteration}
  \Require $\mathbf{q}$
  \Ensure $\mathbf{r}$

  \ForAll {$j \in \mathbf{[0,|\mathcal{M}|)}$}
  \ForAll {$i \in \mathbf{[0,|\mathcal{N}|)}$}
      \State $r_{i,j}^0 = 1 $
      \State $r_{i,j}^1 = 1 $
      \ForAll {$j' \in {\mathcal{N}_i  \setminus  j}$}
	    \State $r_{i,j}^0 =r_{i,j}^0 (1 - 2q_{i,j'}^1 ) $
      \EndFor
     \State $r_{i,j}^0 =1/2 + 1/2r_{i,j}^0  $
     \State $r_{i,j}^1 = 1-r_{i,j}^0$
  \EndFor
  \EndFor

\EndProcedure

\Statex

\Statex

\end{algorithmic}
\label{alg:MessagePass}
\end{algorithm}
\vspace{-3mm}

\begin{algorithm}[h!]
\renewcommand{\algorithmicrequire}{\textbf{\quad\,\, Input:}}
\renewcommand{\algorithmicensure}{\textbf{\quad\,\, Output:}}
\caption{Soft-decision decoding}
\begin{algorithmic}[1]

\Procedure {DecodeAWGN}{}
\Comment{SP algorithm}
  \Require $\mathbf{y}$ -- output from a demodulator \newline $\mathrm{ITERATIONS}$ -- maximum number of iterations
      \newline $\sigma$ -- variance of the channel
  \Ensure $\widehat{\mathbf{c}}$
  \State $\mathbf{q} = $Initialize$(\mathbf{p}, \sigma)$	
  \Comment{See Algorithm~\ref{alg:CalcInit}}
  \State $\mathbf{r} = $Values to Variable Nodes$(\mathbf{q})$	
  \newline
  \Comment{See Algorithm~\ref{alg:MessagePass} (serial) or~\ref{alg:MessagePassPar} (parallel approach)}
  \State $\widehat{\mathbf{c}} = $Calculate Estimation$(\mathbf{r})$
	  \Comment{See Algorithm~\ref{alg:MessagePass} (serial) or~\ref{alg:MessagePassPar} (parallel approach)}
   \If{$\widehat{\mathbf{c}} \mathbf{H}^T = \mathbf{0} $ }
   \quad     \Return $\widehat{\mathbf{c}}$    
    \EndIf	
  \For {$it \in (0,\mathrm{ITERATIONS} ) $}
	  \State $\mathbf{q} = $Values to Check Nodes$(\mathbf{r})$	
	  \Comment{See Algorithm~\ref{alg:CalcEstimation} (serial) or~\ref{alg:CalcEstimationPar} (parallel approach)}
	  \State $\mathbf{r} = $Values to Variable Nodes$(\mathbf{q})$	
	  \Comment{See Algorithm~\ref{alg:MessagePass} (serial) or~\ref{alg:MessagePassPar} (parallel approach)}
	  \State $\widehat{\mathbf{c}} = $Calculate Estimation$(\mathbf{r})$
	  \If{$\widehat{\mathbf{c}} \mathbf{H}^T = \mathbf{0} $ }
	   \quad     \Return $\widehat{\mathbf{c}}$    
	  \Comment{See Algorithm~\ref{alg:CalcSyndromePar} for parallel approach}
	    \EndIf
  \EndFor
\EndProcedure
\Statex

\end{algorithmic}
\label{alg:decode}
\end{algorithm}

\begin{algorithm}[h!]
\renewcommand{\algorithmicrequire}{\textbf{\quad\,\, Input:}}
\renewcommand{\algorithmicensure}{\textbf{\quad\,\, Output:}}
\caption{Initialize step}
\begin{algorithmic}[1]

\Procedure {Initialize }{}
\Comment{Probabilities for AWGN}
  \Require $\mathbf{y}$, $\sigma$
  \Ensure $\mathbf{q}$

  \ForAll {$y_j \in \mathbf{y}$}
    \State $p_j =1.0/(1 +  \mathrm{exp}(-2y_j / \sigma^2) )$
  \EndFor
  \ForAll {$j \in \mathbf{[0,|\mathcal{M}|)}$}
  \ForAll {$i \in \mathbf{[0,|\mathcal{N}|)}$}
    \State $q_{i,j} =p_j$
  \EndFor
  \EndFor


\EndProcedure

\Statex

\end{algorithmic}
\label{alg:CalcInit}
\end{algorithm}

\begin{algorithm}[h!]
\renewcommand{\algorithmicrequire}{\textbf{\quad\,\, Input:}}
\renewcommand{\algorithmicensure}{\textbf{\quad\,\, Output:}}
\caption{Calculation of the estimation}
\begin{algorithmic}[1]

\Procedure {Calculate Estimation}{}
\newline
\Comment{See Algorithm~\ref{alg:MessagePassPar} for parralel approach}
  \Require $\mathbf{p}$, $\mathbf{r}$
  \Ensure $\mathbf{\widehat{c}}$

  \ForAll {$j \in \mathbf{[0,|\mathcal{M}|)}$}

      \State $Q_{i,j}^0 =p_j^0 $
      \State $Q_{i,j}^1 =p_j^1 $
      \ForAll {$i \in {\mathcal{M}_j  }$}
	    \State $Q_{i,j}^0 =Q_{i,j}^0 r_{i,j}^0 $
	    \State $Q_{i,j}^1 =Q_{i,j}^1 r_{i,j}^1 $
      \EndFor

    \If{$Q_{i,j}^0>Q_{i,j}^1$}
      \quad ${\mathrm{\widehat{c}}_j}=0$
    \Else
      \quad ${\mathrm{\widehat{c}}_j}=1$
    \EndIf

  \EndFor

\EndProcedure

\Statex

\end{algorithmic}
\label{alg:CalcEstimation}
\end{algorithm}


\section{Parallelization of LDPC decoding using GPU}
\subsection{Introduction}
The SP algorithm works as an iterative process of message passing between the two sets of nodes (variable and check) in the Tanner graph. Although the number of operations needed to be performed grows with the number of edges in the graph, the algorithm can be accelerated when deployed on massive parallel architectures. Moreover, the potential acceleration achieved by the parallelization of calculations grows with the number of edges in the graph, because more values can be calculated simultaneously. This can lead to interesting applications for long block length codes providing excellent error correcting capabilities.

In recent years, there has been an increasing interest in implementing LDPC decoders in a wide variety of hardware architectures, including GPU. Several contributions deal with fitting the decoder on parallel architectures with the use OpenCL or CUDA frameworks and discuss the benchmarks \cite{gpu_1,gpu_2,gpu_3,gpu_4,gpu_5,gpu_7,gpu_8,gpu_9,gpu_10,gpu_11}. However, work reviewed so far deal mostly with some families of LDPC codes and the application of parallel decoders is limited. In this article, we propose a general parallel approach for the decoder of any irregular LDPC code. The proposed approach divides calculations into a scalable number of threads. Each thread performs the calculation of the value outgoing through the edge, which is associated with the thread itself (edge-level parallelization). The approach was chosen because of its suitability for any irregular LDPC matrices, scalability for any code block lengths and deployablity on many hardware architectures. It is also convenient for derived algorithms for LDPC decoding, such as Min-Sum (MS) or adaptive MS \cite{adaptivems_2010}. In the previous work dealing with the parallel LDPC decoding, the calculations are mostly divided on the level of rows and columns of the parity-check matrices. 

\subsection{Our approach}
In this section, we describe the approach of the edge-level parallelization used for the LDPC decoder. The principle is also shown in the illustrated example supported by consistent figures associated with the same LDPC (14,7) code.
Considering the code given by the parity-check matrix (Fig.~\ref{fig:ParCheckMatrix}) and associated Tanner graph (Fig. 4), we define the following arrays used as address iterators for the parallel message passing algorithm (described in Algorithms 3 and 4):

\begin{itemize}
\item a sorted tuple of variable nodes $\mathbf{v}=(v_j)$ starting with the lowest index and associated tuple of check nodes  $\mathbf{c}=(c_i)$, such $i,j : \mathbf{H}_{i,j}=1$ and $i \in [0, n-k),  j \in [0, n);$ then, $(c_i,v_j)$ unequivocally defines an edge in the Tanner graph; $n$ is the number of variable nodes and $n-k$ is the number of check nodes
\item a tuple of edges $\mathbf{e}=(e_k)=(0,1,2,...,|\mathbf{c}|)$
\item a tuple of connected edges $\mathbf{t}=(t_k)$ with a variable node $v_k$; then, $t_k = |(v_k)|, v_k \in \mathbf{v}$
\item a tuple of starting positions  $\mathbf{s}=(s_k)$ for iterating in order to calculate the value passed through the edge $e_k$;  $s_k=\mathrm{arg~min_k} (v_k : v_k \in \mathbf{v})$
\item a tuple $\mathbf{u}=(u_k)$ of relative positions of the $e_k$ associated with the connected node $v_k$; $u_k = k -|(v_q) : q < k, v_q\neq v_k|$

\end{itemize}

The arrays defined above are used as address iterators for calculations of messages outgoing from variable nodes to check nodes (the first half of the iteration).
We also show the arrays in the illustrative example. Supposing the code (14,7) given by the parity-check matrix in Fig.~\ref{fig:ParCheckMatrix}, the arrays derived by the principle described above are shown in Table~\ref{table:variable_nodes}.
The first half of the iteration of the LDPC decoding process calculates the values passed from the variable nodes to the check nodes. With the use of the array iterators we can perform such calculations without any complicated operations with array indices. The pseudo code is shown in Algorithm~\ref{alg:MessagePassPar}. The local index of the thread  (according to the OpenCL terminology) is denoted as $lid$ and the number of synchronized threads working in parallel is denoted as $lgsize$. 
Because all threads performing the calculations have to be synchronized after they finish writing in the memory and the number of synchronizable threads is strictly limited (e. g. 1024), the calculations are divided in several steps (pages) if necessary. This is when the number of edges is greater than the $lgsize$ variable. An illustrative example for 12 synchronizable threads is shown in~Fig.~\ref{fig:ParCheckMatrix}.

The arrays used for the messages outgoing from the check nodes to the variable nodes are derived similarly. Keeping of the unique edge identifier $(c_i ,v_j)$ and associated edge index $e_k$, the arrays $\mathbf{c}, \mathbf{v}, \mathbf{e}$ are sorted starting with the lowest check node index and other arrays are derived considering the messages outgoing from the check nodes. Such arrays are then denoted as $\overline{\mathbf{e}},  \overline{\mathbf{c}}, \overline{\mathbf{v}}, \overline{\mathbf{t}}, \overline{\mathbf{s}}, \overline{\mathbf{u}}$) in the following descriptions. As a demonstrative example, the arrays for the second half of the iteration are shown in Table~\ref{table:check_nodes}. 

The algorithm performing the second half of the iteration processes the arrays described above. Its pseudo code is shown in Algorithm~\ref{alg:MessagePassPar}. After finishing the second half of the iteration we can continue with the next iteration. The whole decoding principle remains the same, as described in Algorithm~\ref{alg:MessagePass}. 

For example, the address iterators for the LDPC (14,7) code are listed in Table~\ref{table:variable_nodes} and Table~\ref{table:check_nodes}. Both tables are particularly useful for understanding the principle and checking the correctness of the implementation. To keep the consistency and for tutorial purposes, both tables are associated with the LDPC (14,7) code given by the parity-check matrix from Fig.~\ref{fig:ParCheckMatrix}.


\newcolumntype{C}[1]{>{\centering\let\newline\\\arraybackslash\hspace{0pt}}m{#1}}

\begin{table*}[ht]
\scriptsize
\begin{center}
\caption{{\color{black} Addresses used for message calculation outgoing from variable nodes.}}
\label{table:variable_nodes}
\begin{tabular}{|C{1cm}|C{0.5mm}C{0.5mm}C{0.5mm}C{0.5mm}C{0.5mm}C{0.5mm}C{0.5mm}C{0.5mm}C{0.5mm}C{0.5mm}C{0.5mm}C{0.5mm}C{0.5mm}C{0.5mm}C{0.5mm}C{0.5mm}C{0.5mm}C{0.5mm}C{0.5mm}C{0.5mm}C{0.5mm}C{0.5mm}C{0.5mm}C{0.5mm}C{0.5mm}C{0.5mm}C{0.5mm}C{0.5mm}C{0.5mm}C{0.5mm}C{1mm}|}
\hline
\textbf{array} & \multicolumn{31}{c}{\textbf{values}}\\
\hline
\small$\mathbf{e}\;$ & 0 & 1 & 2 & 3 & 4 & 5 & 6 & 7 & 8 & 9 & 10 & 11 & 12 & 13 & 14 & 15 & 16 & 17 & 18 & 19 & 20 & 21 & 22 & 23 & 24 & 25 & 26 & 27 & 28 & 29 & 30 \\
\hline
\small$\mathbf{v}\;$ & 0 & 0 & 0 & 0 & 1 & 1 & 2 & 2 & 3 & 3 & 3 & 4 & 4 & 5 & 5 & 6 & 6 & 7 & 7 & 8 & 8 & 9 & 9 & 10 & 10 & 11 & 11 & 12 & 12 & 13 & 13 \\

\hline
\small$\mathbf{c}\;$ & 5 & 3 & 2 & 0 & 4 & 0 & 5 & 1 & 6 & 4 & 1 & 4 & 3 & 1 & 0 & 4 & 2 & 6 & 5 & 5 & 4 & 2 & 1 & 6 & 0 & 3 & 1 & 6 & 3 & 5 & 0 \\

\hline
\small$\mathbf{t}\;$ & 4 & 4 & 4 & 4 & 2 & 2 & 2 & 2 & 3 & 3 & 3 & 2 & 2 & 2 & 2 & 2 & 2 & 2 & 2 & 2 & 2 & 2 & 2 & 2 & 2 & 2 & 2 & 2 & 2 & 2 & 2 \\

\hline
\small$\mathbf{s}\;$ & 0 & 0 & 0 & 0 & 4 & 4 & 6 & 6 & 8 & 8 & 8 & 11 & 11 & 13 & 13 & 15 & 15 & 17 & 17 & 19 & 19 & 21 & 21 & 23 & 23 & 25 & 25 & 27 & 27 & 29 & 29 \\

\hline
\small$\mathbf{u}\;$ & 0 & 1 & 2 & 3 & 0 & 1 & 0 & 1 & 0 & 1 & 2 & 0 & 1 & 0 & 1 & 0 & 1 & 0 & 1 & 0 & 1 & 0 & 1 & 0 & 1 & 0 & 1 & 0 & 1 & 0 & 1 \\
\hline
\end{tabular}
\end{center}
\end{table*}
\begin{table*}[ht]
\scriptsize
\begin{center}
\caption{{\color{black} Addresses used for message calculation outgoing from check nodes.}}
\label{table:check_nodes}
\begin{tabular}{|C{1cm}|C{0.5mm}C{0.5mm}C{0.5mm}C{0.5mm}C{0.5mm}C{0.5mm}C{0.5mm}C{0.5mm}C{0.5mm}C{0.5mm}C{0.5mm}C{0.5mm}C{0.5mm}C{0.5mm}C{0.5mm}C{0.5mm}C{0.5mm}C{0.5mm}C{0.5mm}C{0.5mm}C{0.5mm}C{0.5mm}C{0.5mm}C{0.5mm}C{0.5mm}C{0.5mm}C{0.5mm}C{0.5mm}C{0.5mm}C{0.5mm}C{1mm}|}
\hline
\textbf{array} & \multicolumn{31}{c}{\textbf{values}}\\
\hline
\small$\mathbf{\overline{e}}\;$ & 3 & 5 & 14 & 24 & 30 & 7 & 10 & 13 & 22 & 26 & 2 & 16 & 21 & 1 & 12 & 25 & 28 & 4 & 9 & 11 & 15 & 20 & 0 & 6 & 18 & 19 & 29 & 8 & 17 & 23 & 27 \\
\hline
\small$\mathbf{\overline{v}}\;$ & 0 & 1 & 5 & 10 & 13 & 2 & 3 & 5 & 9 & 11 & 0 & 6 & 9 & 0 & 4 & 11 & 12 & 1 & 3 & 4 & 6 & 8 & 0 & 2 & 7 & 8 & 13 & 3 & 7 & 10 & 12  \\
\hline
\small$\mathbf{\overline{c}}\;$ & 0 & 0 & 0 & 0 & 0 & 1 & 1 & 1 & 1 & 1 & 2 & 2 & 2 & 3 & 3 & 3 & 3 & 4 & 4 & 4 & 4 & 4 & 5 & 5 & 5 & 5 & 5 & 6 & 6 & 6 & 6 \\
\hline
\small$\mathbf{\overline{t}}\;$ & 5 & 5 & 5 & 5 & 5 & 5 & 5 & 5 & 5 & 5 & 3 & 3 & 3 & 4 & 4 & 4 & 4 & 5 & 5 & 5 & 5 & 5 & 5 & 5 & 5 & 5 & 5 & 4 & 4 & 4 & 4 \\
\hline
\small$\mathbf{\overline{s}}\;$ & 0 & 0 & 0 & 0 & 0 & 5 & 5 & 5 & 5 & 5 & 10 & 10 & 10 & 13 & 13 & 13 & 13 & 17 & 17 & 17 & 17 & 17 & 22 & 22 & 22 & 22 & 22 & 27 & 27 & 27 & 27 \\
\hline
\small$\mathbf{\overline{u}}\;$ & 0 & 1 & 2 & 3 & 4 & 0 & 1 & 2 & 3 & 4 & 0 & 1 & 2 & 0 & 1 & 2 & 3 & 0 & 1 & 2 & 3 & 4 & 0 & 1 & 2 & 3 & 4 & 0 & 1 & 2 & 3  \\
\hline
\end{tabular}
\end{center}
\end{table*}




\begin{algorithm}[h]
\renewcommand{\algorithmicrequire}{\textbf{\quad\,\, Input:}}
\renewcommand{\algorithmicensure}{\textbf{\quad\,\, Output:}}
\caption{Parallel message passing}
\begin{algorithmic}[1]

\Procedure {Iterate to Check Nodes}{}
\newline
\Comment{Half on an iteration}
  \Require $\mathbf{r}$ -- incoming values
     $\mathbf{e}$, $\mathbf{s}$, $ \mathbf{t}$,  $\mathbf{u}$
  \Ensure $\mathbf{q}$

  \For {($p$  = 0; $p<$ totaledges; $p+= lgsize$)}
      \For {$i$  = $s_{lid+p}$ to $s_{lid+p}+t_{lid+p}-1$ }
	   \If{$i = u_{lid+p} +s_{lid+p} $}
	      \quad continue
	    \EndIf
  	      \State $value$ = perform calculations 
	      \Comment{Algorithm~\ref{alg:MessagePass}}
      \EndFor
      \State $index = e_{lid+p}$
      \State $\mathit{q_{index} } = value$
  \EndFor

\EndProcedure

\Statex

\Procedure {Iterate to Variable Nodes}{}
\newline
\Comment{Half on an iteration}
  \Require $\mathbf{q}$ -- incoming values
     $\overline{\mathbf{e}}$, $\overline{\mathbf{s}}$, $\overline{ \mathbf{t}}$,  $\overline{\mathbf{u}}$
  \Ensure $\mathbf{r}$

  \For {($p$  = 0; $p<$ totaledges; $p+=lgsize$)}
      \For {$i$  = $\overline{s_{lid+p}}$ to $\overline{s_{lid+p}} + \overline{t_{lid+p}} - 1$ }
	   \If{$i = \overline{ u_{lid+p} } + \overline{ s_{lid+p} } $} 
	      \quad continue
	    \EndIf
  	      \State $value$ = perform calculations 
	      \Comment{Algorithm~\ref{alg:MessagePass}}
      \EndFor
      \State $index = \overline{e_{lid+p} }$
      \State ${ \mathit{r_{index} } } = value$
  \EndFor

\EndProcedure

\Statex

\end{algorithmic}

\label{alg:MessagePassPar}
\end{algorithm}

\begin{algorithm}[h]
\renewcommand{\algorithmicrequire}{\textbf{\quad\,\, Input:}}
\renewcommand{\algorithmicensure}{\textbf{\quad\,\, Output:}}
\caption{Parallel calculation of the estimation}
\begin{algorithmic}[1]

\Procedure {Calculate Estimation}{}
\newline
\Comment{Parallel approach}
  \Require $\mathbf{r}$ -- incoming values
      $\mathbf{s}$, $ \mathbf{t}$, $\mathbf{v}$
  \Ensure $\mathbf{\widehat{c}}$

  \For {($p$  = 0; $p<$ totaledges; $p+= lgsize$)}
      \State $Q_1$ = $r_{lid+p}$
      \State $Q_0$ = $1 - r_{lid+p}$

      \For {$i$  = $s_{lid+p}$ to $s_{lid+p}+t_{lid+p}-1$ }

	 \State $Q_1$ = $Q_1 r_{i+p}$
            \State $Q_0$ = $Q_0 (1 - r_{i+p} )$

      \EndFor

        \State $index = v_{lid+p} $
        \If{$ Q_1 > Q_0 $} 
	      \quad $\widehat{c}_{index}$ = 1
        \Else
	      \quad $\widehat{c}_{index}$ = 0
        \EndIf

      \State $index = v_{lid+p}$
      \State $\mathit{q_{index} } = value$
  \EndFor

\EndProcedure

\Statex

\end{algorithmic}

\label{alg:CalcEstimationPar}
\end{algorithm}

\begin{algorithm}[h]
\renewcommand{\algorithmicrequire}{\textbf{\quad\,\, Input:}}
\renewcommand{\algorithmicensure}{\textbf{\quad\,\, Output:}}
\caption{Parallel calculation of the syndrome}
\begin{algorithmic}[1]

\Procedure {Calculate Syndrome}{}
\newline
\Comment{Parallel approach}
 \Require $\mathbf{\widehat{c}}$ -- codeword estimation,
    $\overline{\mathbf{s}}$, $\overline{ \mathbf{t}}$,  $\overline{\mathbf{c}}$, $\overline{\mathbf{v}}$
  \Ensure $\mathbf{z}$ -- syndrome $\widehat{\mathbf{c}} \mathbf{H}^T$

  \For {($p$  = 0; $p<$ totaledges; $p+= lgsize$)}
      \State $value$ = 0
      \For {$i$  = $\overline{s_{lid+p}}$ to $\overline{s_{lid+p}}+\overline{t_{lid+p}}-1$ }
           \State $index = \overline{v_{lid+p}} $
	 \State $value$ \^ ~= $\widehat{c}_{index}$

      \EndFor

        \State $index = \overline{c_{lid+p}} $
        \State $z_{index} = value $

  \EndFor

\EndProcedure

\Statex

\end{algorithmic}

\label{alg:CalcSyndromePar}
\end{algorithm}


\section{Bit Error Rate Simulator}
Apart from the implementation of the LDPC decoder, we also considered a Bit Error Rate simulator based on the Additive White Gaussian Noise (AWGN). The simulator is a highly useful tool for benchmarks and code evaluation purposes. The code evaluation requires up to billions of operations to be performed and it is the most time-consuming part of algorithms designing new and innovative LDPC codes. Therefore, its parallelization leads to a significant acceleration of a code design process and more precise simulations become possible. Fast simulations are also needed for evaluating candidate solutions when applying algorithms for performing LDPC code optimizations.

For BER calculation, codewords are modulated and transmitted through the AWGN channel given by the parameter $\sigma$ (often recalculated to the $E_b/N_0$ ratio), as can be seen in Fig.~\ref{fig:ber}. The decoder then receives noised vectors, which are decoded, and Hamming distances between decoded vectors and original codewords are calculated. Due to the linearity of LDPC codes, it is enough to transmit only zero codewords and count the number of 1's at the output of the decoder (Fig.~\ref{fig:ber}). 

\begin{equation}
\sigma ^2 =  \frac{1}{R} \frac{N_0}{2}
\end{equation}	
\begin{equation}
R =  \frac{k}{n}
\end{equation}	
where $k$ is the length of the information message, $n$ is the length of the codeword, $E_b$ is the energy per bit, and $N_0$ is the noise power spectral density.


\section{OpenCL and CUDA implementation}

In current signal and data processing systems, there is an unambiguous trend to use parallel architectures to increase the processing speed, which plays a crucial role in real time applications and determines a deployability of computationally complex algorithms in hardware. Hardware devices supporting massively parallel processing algorithms generally include Graphics Processing Units (GPUs), which are considered in this tutorial article.

In this work, the CUDA and the OpenCL frameworks are used for GPU computations. The OpenCL is an open standard for parallel programming using the different computational devices, such as CPU, GPU, or FPGA. It provides a programming language based on the C99 standard. Unlike OpenCL, CUDA is only for NVIDIA devices starting from G80 series (so called CUDA-enabled GPUs). CUDA gives a possibility to write programs based on the C/C++ and Fortran languages. OpenCL and CUDA programming models are illustrated in Fig. 7.

\subsection{Necessary considerations}

When implementing an algorithm on GPU platform using OpenCL or CUDA frameworks, two main issues have to be considered:

\begin{itemize} 
 \item size of the local memory (OpenCL) or shared memory (CUDA),
 \item size of the working group (OpenCL) or block size (CUDA).
\end{itemize}

\begin{figure*}[h!]
    \centering
    \begin{subfigure}[h]{0.37\textwidth}
        \centering
        \includegraphics[width=\textwidth]{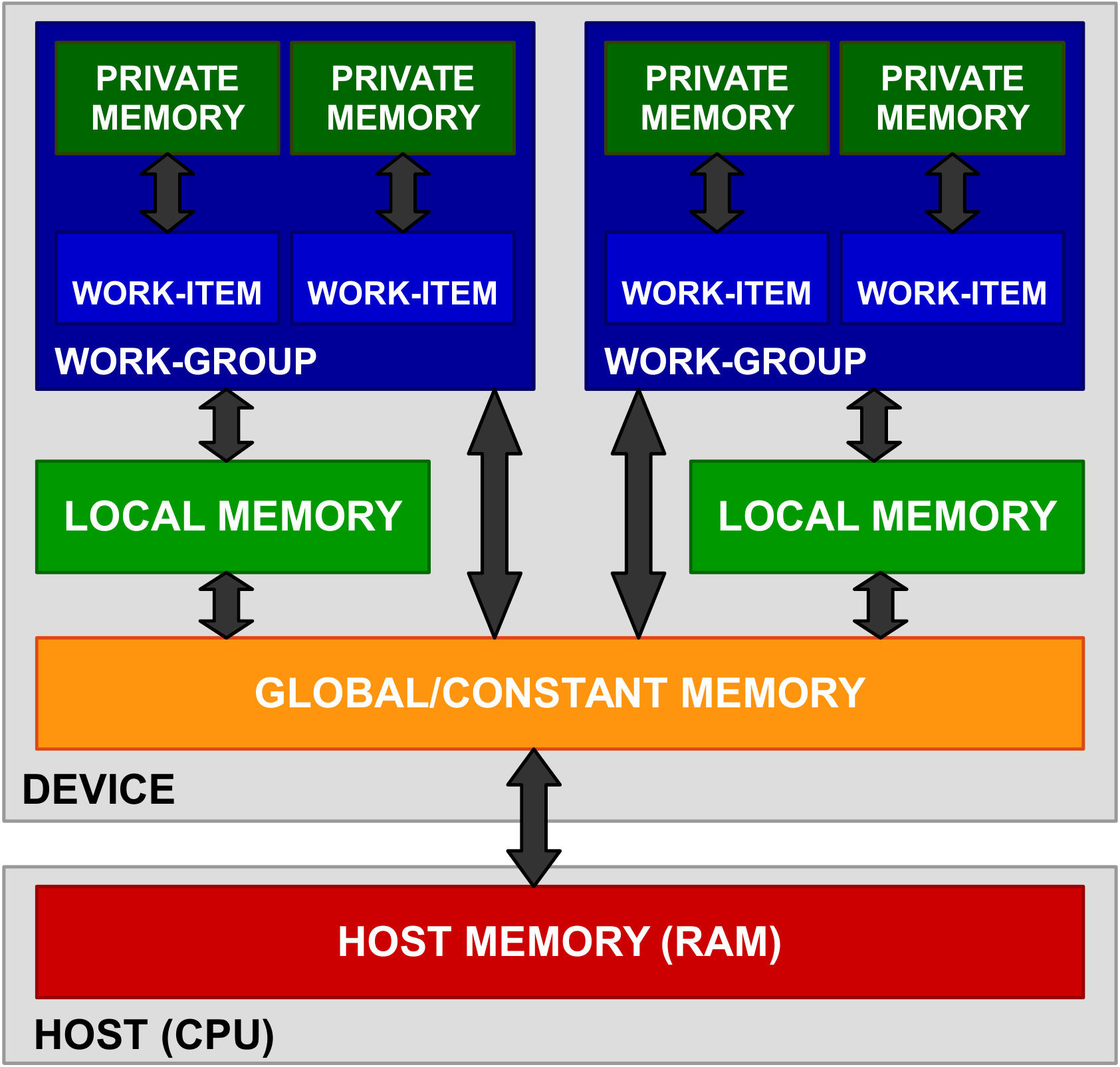}
        \caption{OpenCL programing model.}
        \label{fig:ocl}
    \end{subfigure}
    ~
    \begin{subfigure}[h]{0.37\textwidth}
        \centering
        \includegraphics[width=\textwidth]{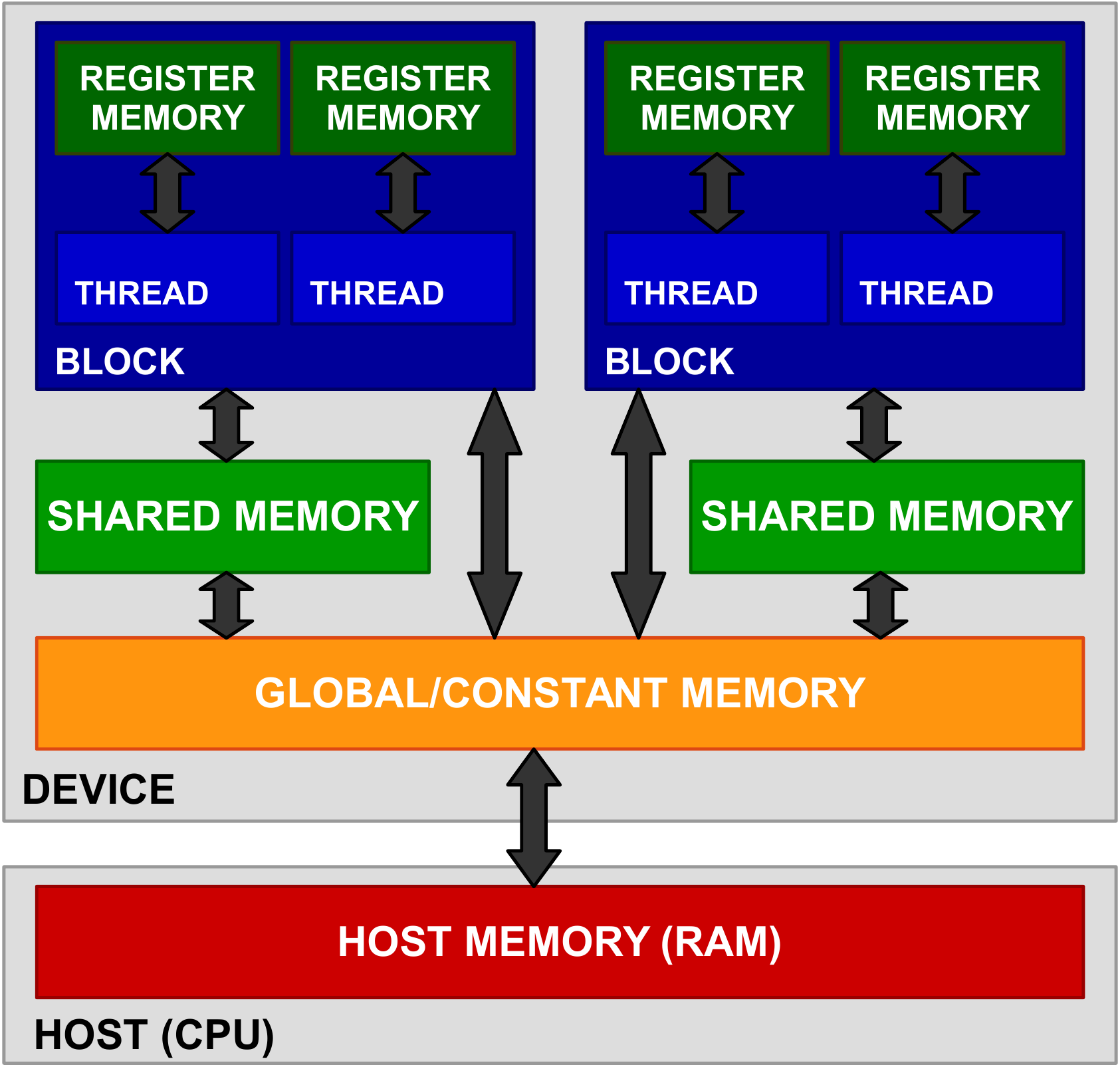}
        \caption{CUDA programming model.}
        \label{fig:cuda}
    \end{subfigure}
    \caption{OpenCL and CUDA programming models.}
    \label{fig:ocl_and_cuda}
\end{figure*}

GPU devices offer several types of the allocable memory, which differ in their speed and their size. The memory type used to store variables is specified in the source code by the prefix according to the OpenCL or CUDA syntax rules. Generally, the largest allocable size, typically in gigabytes for current devices, is located in the global memory. However, the global memory is also the slowest one. A higher speed is provided by the local memory, but the size is typically only in kilobytes. Exceeding the limited size of the local memory usually leads to incorrect results without any warnings in the compilation report.

Another crucial issue related to an algorithm implementation in GPU devices is the working group size. Although the GPU can run thousands of threads in parallel, these threads are not synchronized among each other in terms of writing in the memory. The threads are split into working groups and they can be synchronized only among other threads at the same working group. The size of the working groups is strictly limited (typically 1024).

\subsection{Coding}

Both frameworks processes two types of code
\begin{itemize} 
 \item host (runtime), running serially on CPU
 \item kernel (device), running parallely on GPU
\end{itemize}

\newbox\MyFirstListing
\newbox\tabbox

\begin{listing}[ht]
\setbox\MyFirstListing\hbox to 0.85\linewidth{\vbox{\setbox\tabbox\hbox{\setlength\tabcolsep{0pt}\begin{tabular}[t]{p{0.3\textwidth}} Types \end{tabular}} \caption{\usebox\tabbox\hspace*{0.05\linewidth}} \label{lst:types}

\quad\begin{sublisting}[t]{0.85\linewidth}

\lstset{
    basicstyle=\scriptsize,
}

\begin{lstlisting}
typedef struct Edge{
   int index;	// e array
   int vn;		// v array
   int cn;		// c array
   int edgesConnectedToNode;	// t array
   int absoluteStartIndex;		// s array
   int relativeIndexFromNode;	// u array
} Edge;

typedef struct EdgeData{
   double passedValue;
} EdgeData;

typedef struct CodeInfo{
   int totalEdges;	// number of edges
   int varNodes;		// number of variable nodes
   int checkNodes;	// number of check nodes
} CodeInfo;
\end{lstlisting}
\end{sublisting}}}

\centering
\fcolorbox{black}{background}{\usebox\MyFirstListing}
\end{listing}

%
%

The kernel is executed by the host. In CUDA, the kernel execution is more straightforward compared to OpenCL, as can be seen in the consistent examples in Listing~\ref{lst:host}. Both codes execute the kernel $berSimulate$ in 100 working groups (blocks) with 512 threads per one working group. After finishing the kernel, the results are copied in the $berOut$ array and processed by the host.
Because the kernel function has to be considered as a function running in parallel, each thread has its own unique identifier - the combination of global ID and local ID in OpenCL or the combination of thread ID and block ID in CUDA, which can be recalculated vice versa. 
The parallel implementation of the function $decodeAWGN$, defined in Algorithm~\ref{alg:decode}, is shown in Listing~\ref{lst:decoder}. Types used for code definition and passing messages are pointed in Listing~\ref{lst:types}.

Some main differencies between the OpenCL and CUDA syntax rules are shown in Table~\ref{table:ocl_and_cuda_syntax}, which can be used when moving the source code from one framework to another one. 


\begin{table*}[h!]
\begin{center}
\caption{\color{black} Comparison of chosen OpenCL and CUDA syntax rules}
\label{table:ocl_and_cuda_syntax}
\begin{tabular}{c  |  c | c }
\hline
	\rule{0pt}{3ex}  
\textbf{command} &  \textbf{OpenCL} &  \textbf{CUDA} \;\;  \\
 
\hline
	\rule{0pt}{3ex}  
\small thread synchronization & \small \texttt{barrier(CLK\_GLOBAL\_MEM\_FENCE);}   & \small \texttt{\_\_syncthreads();}  \\
	\rule{0pt}{3ex}  
\small kernel prefix & \small  \texttt{\_\_kernel}   & \small \texttt{\_\_global\_\_}  \\
	\rule{0pt}{3ex}  
\small local memory prefix & \small  \texttt{\_\_local}   & \small \texttt{\_\_shared\_\_}  \\
	\rule{0pt}{3ex}  
\small get local ID & \small  \texttt{int lid = get\_local\_id(0);}   & \small \texttt{int lid = threadIdx.x;}  \\
	\rule{0pt}{3ex}  
\small get global ID & \small  \texttt{int gid = get\_global\_id(0);}   & \small \texttt{int gid = blockIdx.x} \\
	\rule{0pt}{3ex}  
 &    & \small \texttt{* blockDim.x+ threadIdx.x;} \\

\hline



\end{tabular}
\end{center}
\end{table*}



\section{Results}
\subsection{Experimental evaluation}
Developed algorithms for LDPC decoding were run on NVIDIA Tesla K40 (Atlas) and Intel Xeon E5-2695v2 platforms \cite{smith_AT_2014, nvidia_gk210_2014}. The NVIDIA device contains 2880 CUDA cores and runs at 745 MHz. The peak performance for double precision computations with floating point is 1.43 Tflops. The clock frequency of the Intel Xeon CPU is 2.4 GHz. All measurements include the time required for random generation, realised by the Xorshift+ algorithm and the Box-Muller transform.

Benchmarks were performed through the calculation of the Bit Error Rate at $E_b/N_0 = 2 \mathrm{dB}$ for a code given by the NASA CCSDS standard \cite{nasa_256} and its protographically expanded derivations \cite{thorpe_2003}, \cite{fang_2015}. Based on the results obtained from NVIDIA Tesla K80, we got slightly better performance with the use of the CUDA framework, as shown in Fig. 9. Compared to the CPU implementation run on Intel Xeon, the acceleration grows with the size of working groups and the number of decoders running in parallel to the limit of the device, as illustrated in Fig. 8.
GPU become very effective for longer block length codes, as also shown in Table~\ref{table:ocl_and_cuda_results}. The ratio between CPU (C++ compiler with O3 optimization) and GPU was 25 for code of 262144 bits.

\subsection{Further acceleration}%
To keep the generality, no simplifications in the decoding algorithm were applied and the experimental evaluation was performed with the use of the global memory. 
For further acceleration, several tasks can be considered, i. e. usage of the local memory, variables with a lower precision, look-up tables, or modifications of the algorithm for certain families of LDPC codes. 
For example, by moving the part of variables in the local (shared) memory, the decoder works approximately 40\% faster in our experience. However, it is not possible to decode longer codewords because of the size limitations (240 kB of the local memory per working group).
Another possibility for greater optimization could be the parallelization of less computationally intensive functions. After applying parrallel algorithms for passing messages, calculating the syndrome and the estimation, the most serial time-consuming operation is checking syndrome for all zero equality (approximately 34\% of the decoding function in our experience). 



\begin{figure*}[h!]
    \centering
    \begin{subfigure}[b]{0.4\textwidth}
        \centering
        \includegraphics[width=\textwidth]{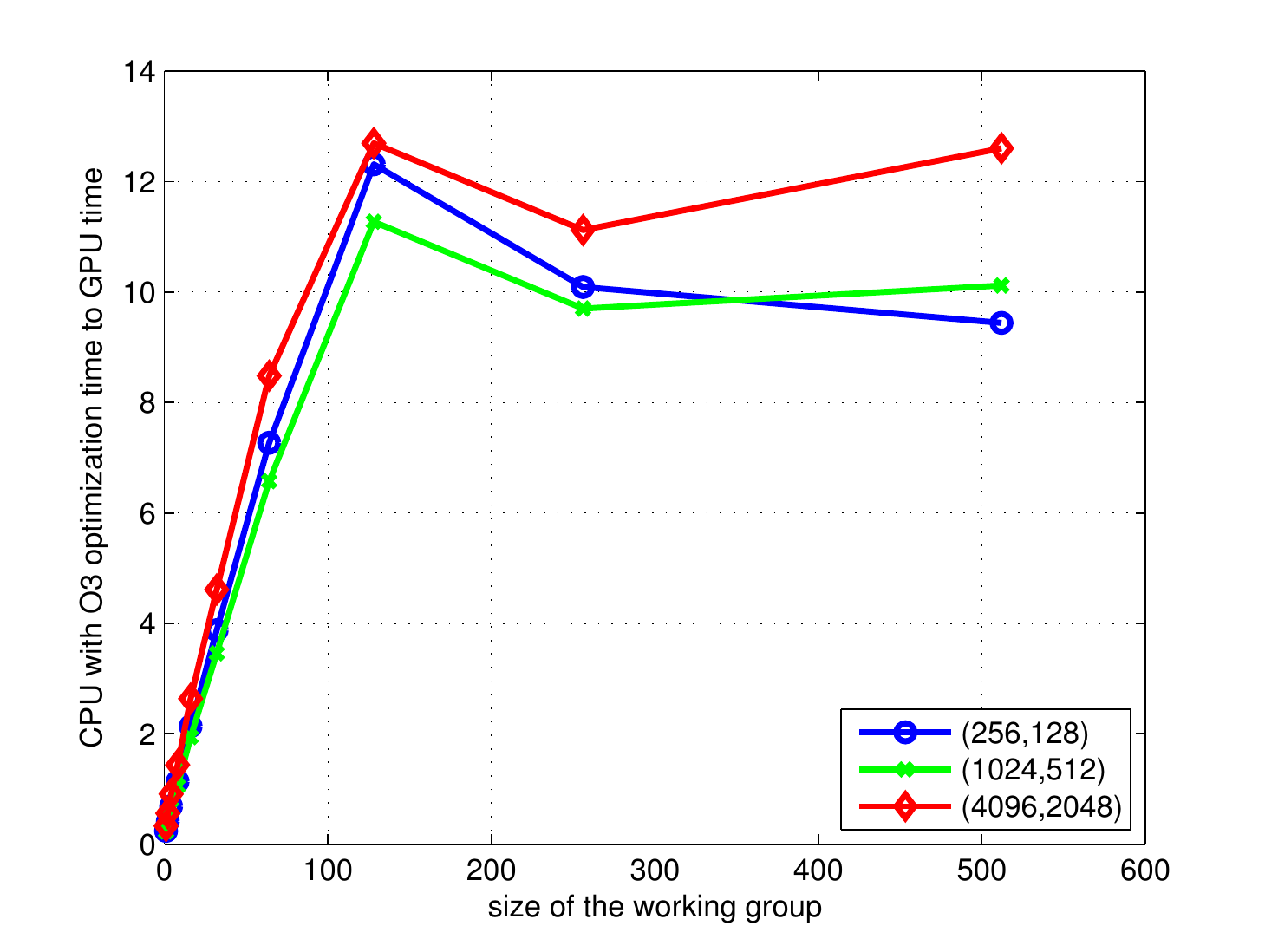}
        \caption{Acceleration dependence on the block (working group) for 100 decoders running in parallel.\\}
        \label{fig:acceleration1}
    \end{subfigure}
    ~
    \begin{subfigure}[b]{0.4\textwidth}
        \centering
        \includegraphics[width=\textwidth]{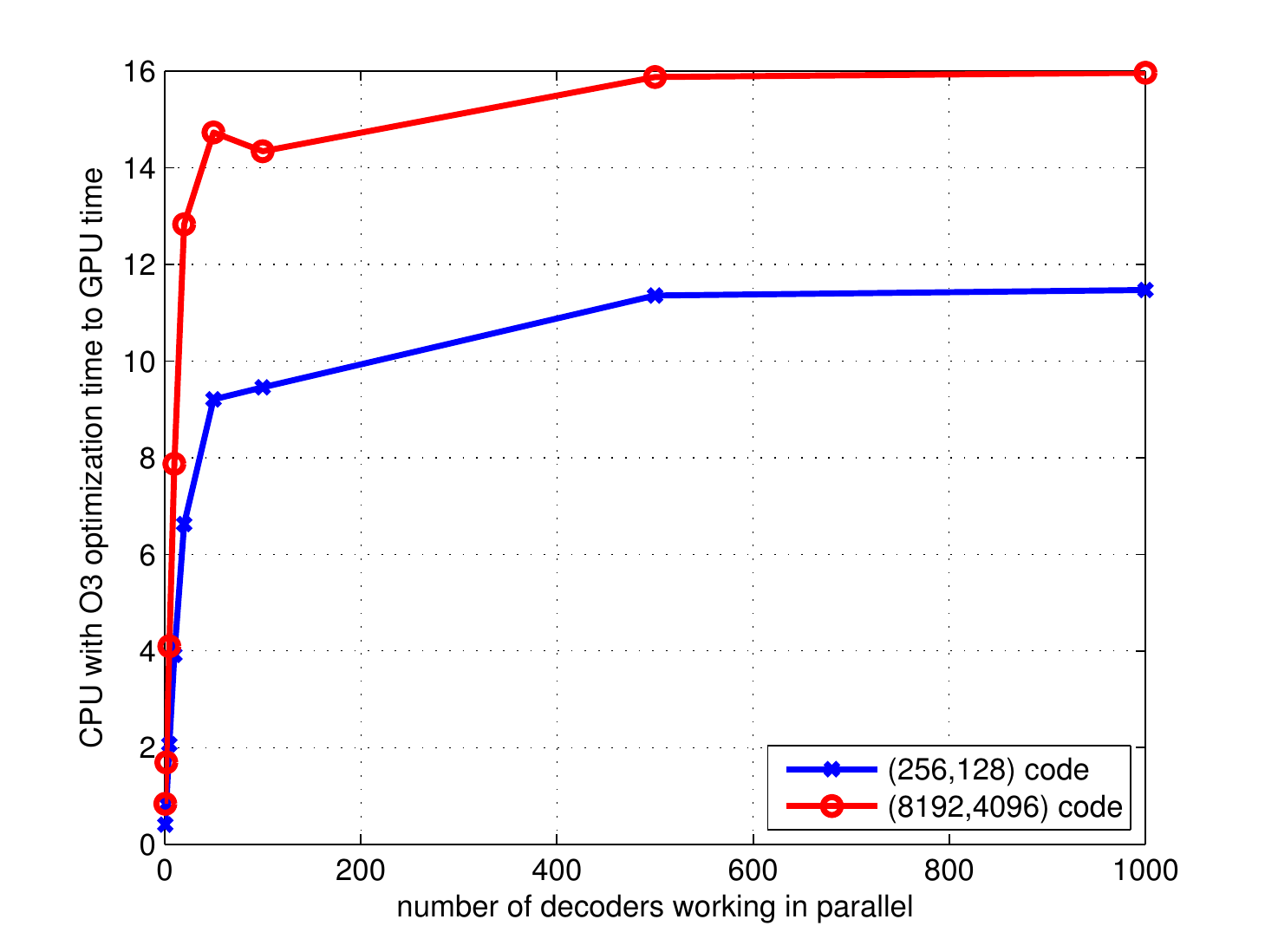}
        \caption{Acceleration dependence on the number of decoders working in parallel when the size of the working group is 512.}
        \label{fig:acceleration2}
    \end{subfigure}
    \caption{Measured acceleration with the use of the CUDA framework.}
    \label{fig:acceleration}
\end{figure*}

\begin{figure}[h]
    \centering
        \includegraphics[width=0.5\textwidth]{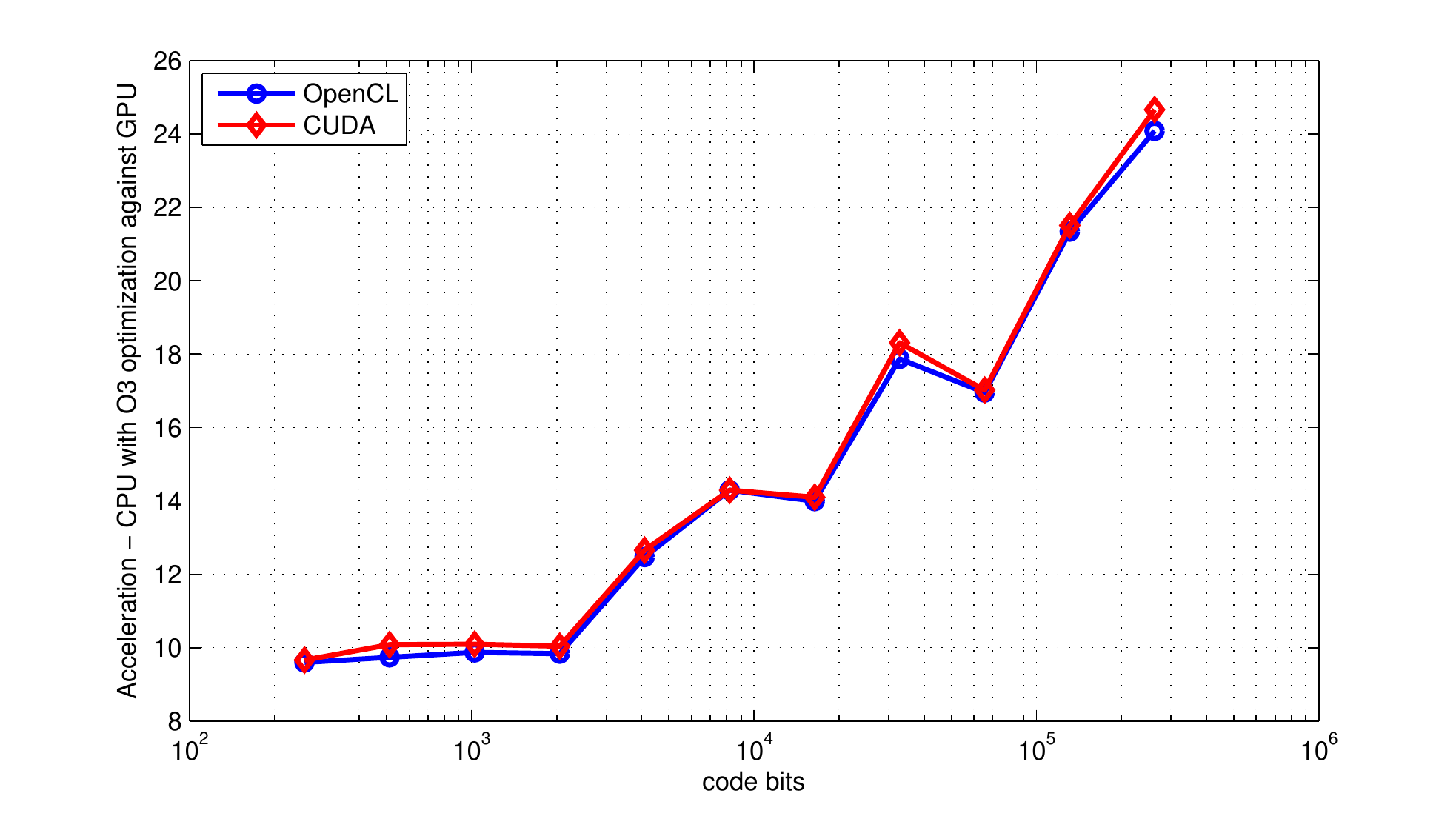}
        \caption{Acceleration dependence on the length of the code. Comparison for OpenCL and CUDA frameworks (local group of 512 threads and 100 decoders working in parallel) against CPU implementation using C++ compiler with O3 optimization. Time was mesuared for 10000 decoded codewords at $E_b/N_0=$2dB.}
        \label{fig:AccComparison}
\end{figure}

\begin{table*}[h]
\small
\begin{center}
\caption{{\color{black} Comparison for OpenCL and CUDA framework (local group of 512 threads and 100 decoders working in parallel) against the CPU implementation using C++ compiler with O3 optimization. Time was mesuared for 10000 decoded codewords at $E_b/N_0=$2dB.}}
\label{table:ocl_and_cuda_results}
\begin{tabular}{  c | c| c |c|c|c }
\hline
	\rule{-4pt}{3ex}  
\textbf{code} &  \textbf{edges} \;\; & \textbf{OpenCL} &   \;\;\,\textbf{CUDA}  &  \;\,\textbf{C++}  &  \;\,\textbf{C++ with O3 optimization} \\
\hline
	\rule{0pt}{3ex}  
 (256,128) & 1024 & 0.32 s & 0.32 s & 24.24 s & 3.11 s \\
 (512,256) & 2048 & 0.64 s  & 0.61 s  &26.98 s  & 6.24 s \\
 (1024,512) & 4096 & 1.26 s  & 1.24 s & 99.59 s  & 12.52 s  \\
 (2048,1024) & 8192 & 2.56 s & 2.51 s  & 105.56 s & 25.27 s  \\
 (4096,2048) & 16384 & 5.54 s  & 5.46 s & 415.35 s & 69.17 s \\
 (8192,4096) & 32768 & 12.08 s & 12.08 s & 545.74 s & 172.67 s  \\
 (16384,8192) & 65536 & 26.27 s & 26.08 s & 1717.25 s  & 367.75 s  \\
 (32768,16384) & 131072 & 57.40 s  & 56.02 s  & 2893.91 s  & 1025.9 s  \\
 (65536,32768) & 242144 & 117.31 s & 116.86 s & 8572.08 s & 1989.26 s \\
 (131072,65536) & 524288 & 244.36 s & 242.43 s &  14082.71 s & 5215.11 s \\
 (262144,131072) & 1048576 & 510.06 s & 498.16 s &  35104.28 s & 12287.61 s \\
\hline
\end{tabular}
\end{center}
\end{table*}

\section{Conclusions}
The development of multicore architectures supporting parallel data processing has led to a paradigm shift. Data processing algorithms has to be considered working asynchronously in separated threads, while the threads are synchronized only when writing in registers (memory). Therefore, there is a need for novel approaches and frameworks allowing an algorithm deployabality in modern signal and data processing systems. In this article, we touched with recent frameworks for Graphics Processing Units and probably the best known error correction coding technique, LDPC. In a tutorial-based style, we have provided a general parallel approach for decoding any irregular LDPC code and presented a demonstrative application in consistent examples associated with the LDPC (14,7) code. The presented approaches are based on the edge-level parallelization, where each thread performs the calculation of a particular value passed through the associated edge (one thread for one edge). The potential acceleration achieved by the parallelization of the calculations grows with the number of edges in the graph. This can lead to interesting applications for long block length codes providing excellent error correcting capabilities. 

Hardware devices supporting massively parallel processing algorithms generally include GPUs.  Differencies and similarities, in terms of the terminology and source codes, between the OpenCL and CUDA frameworks used for GPU programming were shown in the paper. Benchmarks for the OpenCL and CUDA approaches were performed on the NASA CCSDS (256,128) standard and its protographically expanded derivations  \cite{thorpe_2003}, \cite{fang_2015}, and the results were compared against the C++ implementation.

Results shown the acceleration which is up to 22 times compared against C++ with O3 optimization, and up to 58 times compared against C++ compilation without optimization. 

Because the OpenCL framework has found utilization in programming FPGA-based systems \cite{fpga_opencl}, the proposed algorithms and their potential modifications can be easily used in a wide variety of fast communication signal processing systems.

\section*{Acknowledgment}

The access to the heterogeneous cluster HybriLIT, provided by the Joint Institute for Nuclear Research, Dubna, Russia, is highly appreciated. 

We would like to thank Vladimir Korenkov and Ivan Stekl for arranging the cooperation, Jan Busa for technical support and especially for professional LaTeX consultations, and Gheorge Adam for his professional comments and interest in this work.

This work was suppported by the project SGS-2015-002 'Modern methods in solution, design and application of electronic and communication systems', by the project of Centre for Advanced Nuclear Technologies, no. TE01020455, and by the JINR grant No. 17-602-01.


\ifCLASSOPTIONcaptionsoff
  \newpage
\fi




\begin{thebibliography}{1}

\bibitem{djordevic_2016}
I. B. Djordjevic, \emph{On Advanced FEC and Coded Modulation for Ultra-High-Speed Optical Transmission}, in IEEE Communications Surveys $\&$ Tutorials, vol. 18, no. 3, pp. 1920-1951, thirdquarter 2016.


\bibitem{hamming_1950}
R. Hamming, \emph{Error detecting and error correcting codes}. Bell Syst. Technical Journal. vol. 29, pp. 41-56, 1950.

\bibitem{spacecraft_2008}
A. Kenneth, D. Divsalar, S. Dolinar, Jon Hamkins, and Fabrizio Pollara. \emph{Design and Standardization of Low-Density Parity-Check Codes for Space Applications}, SpaceOps 2008 Conference, SpaceOps Conferences.

\bibitem{dvb_2009}
ETSI standard, \emph{Digital Video Broadcasting (DVB); Second generation framing structure, channel coding and modulation systems for Broadcasting, Interactive Services, News Gathering and other broadband satellite applications (DVB-S2)}, France, 2009.

\bibitem{ethernet_2007}
10 Gigabit Ethernet: IEEE Standard for Information Technology-Telecommunications and Information Exchange Between Systems-Local and Metropolitan Area Networks-Specific Requirements
Part 3: Carrier Sense Multiple Access With Collision Detection (CSMA/CD) Access Method and Physical Layer Specifications, IEEE Standard 802.3an-2006, Aug. 2006 [Online]. Available: http://standards.ieee.org/getieee802/download/802.3an-2006.pdf


\bibitem{gallager_1963}
R. G. Gallager, \emph{Low Density Parity Check Codes}, Transactions of the IRE Professional Group on Information Theory, Vol. IT-8, January 1962, pp. 2l-28.

\bibitem{bonello_2011}
N. Bonello, S. Chen and L. Hanzo, \emph{Low-Density Parity-Check Codes and Their Rateless Relatives}, in IEEE Communications Surveys $\&$ Tutorials, vol. 13, no. 1, pp. 3-26, First Quarter 2011.

\bibitem{wilberg_1996}
N. Wiberg, \emph{Codes and Decoding on General Graphs}. PhD thesis, Dept. of Electrical
Engineering, Lionköoing, Sweden, 1996. Lionköoing studies in Science and Technologz.
Dissertation No. 440.

\bibitem{opencl_spec}
Khronos OpenCL Working Group, \emph{The OpenCL Specification}, 2011 [Online]. Available: https://www.khronos.org/registry/cl/specs/opencl-1.2.pdf

\bibitem{cuda_spec}
NVIDIA Corporation, \emph{Cuda Runtime API}, Reference manual, 2015 [Online]. Available: http://docs.nvidia.com/cuda/pdf/CUDA\_Runtime\_API.pdf


\bibitem{opencl1}
C. Heinemann, S. S. Chaduvu, A. Byerly and A. Uskov, \emph{OpenCL and CUDA software implementations of encryption/decryption algorithms for IPsec VPNs}, 2016 IEEE International Conference on Electro Information Technology (EIT), Grand Forks, ND, 2016, pp. 0765-0770.

\bibitem{opencl2}
G. Bernabé, G. D. Guerrero and J. Fernández, \emph{CUDA and OpenCL implementations of 3D Fast Wavelet Transform}, Circuits and Systems (LASCAS), 2012 IEEE Third Latin American Symposium on, Playa del Carmen, 2012, pp. 1-4.

\bibitem{opencl3}
J. P. Arun, M. Mishra and S. V. Subramaniam, \emph{Parallel implementation of MOPSO on GPU using OpenCL and CUDA}, 2011 18th International Conference on High Performance Computing, Bangalore, 2011, pp. 1-10.


\bibitem{performance_2011}
J. Fang, A. L. Varbanescu and H. Sips, \emph{A Comprehensive Performance Comparison of CUDA and OpenCL}, 2011 International Conference on Parallel Processing, Taipei City, 2011, pp. 216-225.


\bibitem{gpu_1}
Y. Zhao, X. Chen, C.-W. Sham, Wai M. Tam, and Francis C.M. Lau \emph{Efficient Decoding of QC-LDPC Codes Using GPUs}, Algorithms and Architectures for Parallel Processing. 2011

\bibitem{gpu_2}
G. Falcao, V. Silva, L. Sousa and J. Andrade,  \emph{Portable LDPC Decoding on Multicores Using OpenCL [Applications Corner]}, IEEE Signal Processing Magazine, vol. 29, no. 4, pp. 81-109, July 2012.

\bibitem{gpu_3}
Y. Zhao and F. C. M. Lau, \emph{Implementation of Decoders for LDPC Block Codes and LDPC Convolutional Codes Based on GPUs}, IEEE Transactions on Parallel and Distributed Systems, vol. 25, no. 3, pp. 663-672, March 2014.

\bibitem{gpu_4}
S. Wang, S. Cheng and Q. Wu, \emph{A parallel decoding algorithm of LDPC codes using CUDA}, 2008 42nd Asilomar Conference on Signals, Systems and Computers, Pacific Grove, CA, 2008, pp. 171-175.

\bibitem{gpu_5}
M. Beermann, E. Monró, L. Schmalen and P. Vary, \emph{High speed decoding of non-binary irregular LDPC codes using GPUs}, SiPS 2013 Proceedings, Taipei City, 2013, pp. 36-41.

\bibitem{gpu_7}
Y. Zhao, X. Chen, C.-W. Sham, Wai M. Tam, and Francis C.M. Lau, \emph{Efficient Decoding of QC-LDPC Codes Using GPUs}. 11th International Conference, ICA3PP, Melbourne, Australia, October 24-26, 2011, Proceedings, Part I

\bibitem{gpu_8}
J.-Y. Park and K.-S. Chung, \emph{Parallel LDPC decoding using CUDA and OpenMP}. Park and Chung EURASIP Journal on Wireless Communications and Networking, 2011.

\bibitem{gpu_9}
X. Wen et al., \emph{A high throughput LDPC decoder using a mid-range GPU}, 2014 IEEE International Conference on Acoustics, Speech and Signal Processing (ICASSP), Florence, 2014, pp. 7515-7519.

\bibitem{gpu_10}
G. Wang, M. Wu, B. Yin and J. R. Cavallaro, \emph{High throughput low latency LDPC decoding on GPU for SDR systems}, Global Conference on Signal and Information Processing (GlobalSIP), 2013 IEEE, Austin, TX, 2013, pp. 1258-1261.

\bibitem{gpu_11}
J. Andrade, G. Falcao, V. Silva, \emph{Optimized FastWalsh-Hadamard Transform on GPUs for non-binary LDPC decoding}, Parallel Computing, Vol. 40, 2014, pp. 449–453. 

\bibitem{tanner_1981}
R. M. Tanner, \emph{A Recursive Approach to Low Complexity Codes}. Information Theory, IEEE Transactions, vol.27, no.5, pp.533,547, 1981.

\bibitem{spielman_1998}
D. A. Spielman, \emph{Finding good LDPC codes}, 36th Annual Allerton Conference on Communication, Control, and Computing, 1998.



\bibitem{turbo_1993}
C. Berrou, A. Glavieux, P. Thitimajshima, \emph{Near Shannon limit error-correcting coding and decoding: Turbo-codes},  Communications, 1993. ICC '93 Geneva. Technical Program, Conference Record, IEEE International Conference on , vol.2, no., pp.1064,1070 vol.2, 23-26 May 1993.

\bibitem{rs_1960}
I. Reed, G. Solomon. \emph{Polynomial Codes over Certain Finite Field}. J. Soc. Indust. Appl. Math. vol. 8 pp. 300-304, 1960.


\bibitem{our_1}
J. Broulim, P. Broulim, J. Moldaschl, V. Georgiev and R. Salom, \emph{Fully parallel FPGA decoder for irregular LDPC codes}, Telecommunications Forum Telfor (TELFOR), 2015 23rd, Belgrade, 2015, pp. 309-312.

\bibitem{our_2}
R. Salom and J. Broulim, \emph{LDPC (512,480) genetic design as alternative to CRC in implementation of AODV routing protocol stack}, Telecommunications Forum Telfor (TELFOR), 2015 23rd, Belgrade, 2015, pp. 643-645.

\bibitem{our_3}
J. Broulim and V. Georgiev, \emph{LDPC error correction code utilization}, Telecommunications Forum (TELFOR), 2012 20th, Belgrade, 2012, pp. 1048-1051.

\bibitem{our_4}
J. Broulim, V. Georgiev, J. Moldaschl and L. Palocko, \emph{LDPC code optimization based on Tanner graph mutations}, Telecommunications Forum (TELFOR), 2013 21st, Belgrade, 2013, pp. 389-392.

\bibitem{our_5}
J. Broulim, S. Davarzani, V. Georgiev and J. Zich, \emph{Genetic optimization of a short block length LDPC code accelerated by distributed algorithms}, 2016 24th Telecommunications Forum (TELFOR), Belgrade, 2016, pp. 1-4.


\bibitem{adaptivems_2010}
X. Wu, Y. Song, M. Jiang and C. Zhao, \emph{Adaptive-Normalized/Offset Min-Sum Algorithm}, in IEEE Communications Letters, vol. 14, no. 7, pp. 667-669, July 2010.

\bibitem{smith_AT_2014} Ryan Smith, \emph{NVIDIA Launches Tesla K80, GK210 GPU}. AnandTech (November 17, 2014), \url{http://www.anandtech.com/tag/gpus} (last visit 02/06/2016).

\bibitem{nvidia_gk210_2014} \emph{Whitepaper of NVIDIA’s Next Generation CUDA Compute Architecture: Kepler GK110/210}. \url{http://www.nvidia.com/object/gpu-architecture.html} (last visit 02/06/2016).

\bibitem{nasa_256}
Short Block Length LDPC Codes for TC Synchronization and Channel Coding.  \emph{CCSDS Experimental Specification}. NASA, 2015. 


\bibitem{thorpe_2003}
J. Thorpe, \emph{Low-Density Parity-Check (LDPC) Codes Constructed from Protographs}, IPN Progress Report 42-154, 2003.

\bibitem{fang_2015}
Y. Fang, G. Bi, Y. L. Guan and F. C. M. Lau, \emph{A Survey on Protograph LDPC Codes and Their Applications}, in IEEE Communications Surveys $\&$ Tutorials, vol. 17, no. 4, pp. 1989-2016, Fourthquarter 2015.


\bibitem{fpga_opencl}
\emph{Implementing FPGA Design with the OpenCL Standard}, Altera, 2013.





\end{thebibliography}
%

%


\begin{IEEEbiographynophoto}{Jan Broulim}
Biography text here.
\end{IEEEbiographynophoto}


\begin{IEEEbiographynophoto}{Alexander Ayriyan}
Biography text here.
\end{IEEEbiographynophoto}

\begin{IEEEbiographynophoto}{Vjaceslav Georgiev}
Biography text here.
\end{IEEEbiographynophoto}

\begin{IEEEbiographynophoto}{Hovik Grigorian}
Biography text here.
\end{IEEEbiographynophoto}




\newpage

\onecolumn

\appendix

\linespread{1.0}

\begin{listing}[h!]
\setbox\MyFirstListing\hbox to 0.9\linewidth{\vbox{\vbox{\caption{Execution of the kernel function from runtime\hspace*{0.15\linewidth}} \label{lst:host} \vspace{-0.3cm}}
\vbox{\quad\begin{sublisting}[t]{0.43\linewidth}
\caption{OpenCL source code}\label{lst:host:opencl}
\begin{lstlisting}[linewidth=\textwidth]
// Create an OpenCL context and command queue
// Build the program from source and create kernel
cl_context context=clCreateContext(NULL,CL_DEVICE_TYPE_GPU,&device_id,NULL,NULL,NULL);
cl_cmd_queue cmd_queue = clCreateCommandQueue(context,device_id,0,NULL);
cl_program program=clCreateProgramWithSource(context,1,(const char**)&source,NULL,NULL);
clBuildProgram(program,1,&device_id,NULL,NULL,NULL);
cl_kernel kernelSim = clCreateKernel(program, "berSimulate", NULL);
// Create buffers
cl_mem codeInfo_obj = clCreateBuffer(context, CL_MEM_READ_ONLY,
sizeof(CodeInfo), NULL, NULL);
cl_mem edgesFromVariable_obj =
     clCreateBuffer(context, CL_MEM_READ_ONLY,
     sizeof(Edge) *totalEdges, NULL, NULL);
. . .
// Set kernel parameters
clSetKernelArg(kernelSim, 0, sizeof(cl_mem),
     (void*) &codeInfo_obj);
clSetKernelArg(kernelSim, 1, sizeof(cl_mem),
     (void*) &edgesFromVariable_obj);
. . .
// Copy in the buffers
clEnqueueWriteBuffer(command_queue, codeInfo_obj, CL_TRUE, 0, sizeof(CodeInfo), codeInfo, 0, NULL,NULL);
clEnqueueWriteBuffer(command_queue, edgesFromVariable_obj, CL_TRUE, 0, sizeof(Edge) *totalEdges, edgesFromVariable, 0, NULL, NULL);
int decoders = 100;
local_item_size = 512;
global_item_size = local_item_size * decoders;
// Execute the OpenCL kernel
clEnqueueNDRangeKernel(command_queue, kernelSim, 1, NULL, &global_item_size, &local_item_size, 0, NULL, NULL);
// Copy the results back
clEnqueueReadBuffer(command_queue, berOut_obj, CL_TRUE, 0, sizeof(double) *maxPoints, berOut, 0, NULL, NULL);
\end{lstlisting}
\end{sublisting}\vline\quad
\begin{sublisting}[t]{0.47\linewidth}
\caption{CUDA source code}\label{lst:host:cuda}
\begin{lstlisting}[linewidth=\textwidth]
CodeInfo* codeInfo_obj;
cudaMalloc((void**) &codeInfo_obj, sizeof(CodeInfo));
Edge* edgesFromVariable_obj;
cudaMalloc((void**) &edgesFromVariable_obj,
     sizeof(Edge) *totalEdges);
. . .
// Copy to the device
cudaMemcpy(codeInfo_obj, codeInfo, sizeof(CodeInfo), cudaMemcpyHostToDevice);
cudaMemcpy(edgesFromVariable_obj, edgesFromVariable, sizeof(Edge) *totalEdges,
     cudaMemcpyHostToDevice);
// Same meaning as local_item_size in OpenCL
int block_size = 512;
int decoders = 100;
int blocks = decoders;
// Kernel execution
berSimulate <<< blocks, block_size >>> (codeInfo_obj, edgesFromVariable_obj, edgesFromCheck_obj,
simParams_obj, simulatedPoints_obj, edgeDataInitToCheck_obj, edgeDataToVariable_obj, edgeDataToCheck_obj,
randomGenVariables_obj, estimation_obj, syndrome_obj, noisedVector_obj, berOut_obj);
// Copy the results to host
cudaMemcpy(berOut, berOut_obj,
     sizeof(double) *maxPoints,
     cudaMemcpyDeviceToHost);
\end{lstlisting}
\end{sublisting}}}}

\centering
\fcolorbox{black}{background}{\usebox\MyFirstListing}
\end{listing}

\begin{listing}[h!]
\setbox\MyFirstListing\hbox to 0.9\linewidth{\vbox{\vbox{\caption{Decoder function run from kernel\hspace*{0.15\linewidth}} \label{lst:decoder} \vspace{-0.3cm}}
\vbox{\quad\begin{sublisting}[t]{0.43\linewidth}
\caption{OpenCL source code}\label{lst:decoder:opencl}
\begin{lstlisting}[linewidth=\textwidth]
void decodeAWGN(__global CodeInfo* codeInfo,
__global Edge* edgesFromVariable,
__global Edge* edgesFromCheck,
__global EdgeData* edgeDataInitToCheck,
__global EdgeData* edgeDataToCheck,
__global EdgeData* edgeDataToVariable,
__global int* estimation, 
__global int* syndromeOut,
__global double* noisedVector,
int iterations, double sigma2, int lid,
int totalEdges, int lgsize, int glPageStartIndex)
{
   int index_e, p;
   // initial messages to check nodes
   for (p = 0; p < totalEdges; p += lgsize)
   {
      index_lid = p + lid;
      if (index_lid < totalSize)
         initProbCalcAWGN(noisedVector,
          edgesFromVariable, edgeDataInitToCheck,
          sigma2, glPageStartIndex, index_e);
   }
   barrier(CLK_GLOBAL_MEM_FENCE);
   // iteration back to variable nodes
   for (p = 0; p < totalEdges; p += lgsize)
   {
      index_lid = p + lid;
      if (index_lid < totalSize)
         iterateToVariables(...);
   }
   // calculate the estimation
   barrier(CLK_GLOBAL_MEM_FENCE);
   for (p = 0; p < totalEdges; p += lgsize)
   {
      index_lid = p + lid;
      if (index_lid < totalEdges)
         estimationCalc(...);
   }
   // calculate the syndrome
   barrier(CLK_GLOBAL_MEM_FENCE);
   for (p = 0; p < totalEdges; p += lgsize)
   {
      index_lid = p + lid;
      if (index_lid < totalEdges)
      syndromeCalc(edgesFromCheck, estimation, syndromeOut, glPageStartIndex, index_e);
   }
   barrier(CLK_GLOBAL_MEM_FENCE);   
   int cnodes = codeInfo[0].checkNodes;
   int parity = isAllZero(syndromeOut, cnodes, glPageStartIndex);
   // if syndrome is ok, decoding is successful
   if (parity == 1) return;
   barrier(CLK_GLOBAL_MEM_FENCE);
   int i;
   // forward and back iterations
   for (i = 0; i < iterations; i++)
   {
      // iteration to check nodes
      for (p = 0; p < totalEdges; p += lgsize)
      {
         index_lid = p + lid;
         if (index_lid < totalEdges)
            iterateToCheck(...);
      }
      barrier(CLK_GLOBAL_MEM_FENCE);
      . . .
      // iteration back to variable nodes
      barrier(CLK_GLOBAL_MEM_FENCE);
     . . .
      // calculate the estimation
      barrier(CLK_GLOBAL_MEM_FENCE);
      . . .
      // calculate the syndrome
      barrier(CLK_GLOBAL_MEM_FENCE);
      // if syndrome is ok, decoding is successful
      parity = isAllZero(syndromeOut, cnodes, glPageStartIndex);
      if (parity == 1) return;
   }
}
\end{lstlisting}
\end{sublisting}\vline\quad
\begin{sublisting}[t]{0.47\linewidth}
\caption{CUDA source code}\label{lst:decoder:cuda}
\begin{lstlisting}[linewidth=\textwidth]
__device__ void decodeAWGN(CodeInfo* codeInfo, 
Edge* edgesFromVariable,
Edge* edgesFromCheck, 
EdgeData* edgeDataInitToCheck, 
EdgeData* edgeDataToCheck,
EdgeData* edgeDataToVariable,
int* estimation,
int* syndromeOut,
double* noisedVector,
int iterations, double sigma2, int lid,
int totalEdges, int lgsize, int glPageStartIndex)
{
   int index_e, p;
   // initial messages to check nodes
   for (p = 0; p < totalEdges; p += lgsize)
   {
      index_lid = p + lid;
      if (index_lid < totalSize)
         initProbCalcAWGN(noisedVector, 
          edgesFromVariable, edgeDataInitToCheck,
          sigma2, glPageStartIndex, index_e);
   }
   __syncthreads();
   // iteration back to variable nodes
   for (p = 0; p < totalEdges; p += lgsize)
   {
      index_lid = p + lid;
      if (index_lid < totalSize)
         iterateToVariables(...);
   }
   __syncthreads();
   // calculate the estimation
   for (p = 0; p < totalEdges; p += lgsize)
   {
      index_lid = p + lid;
      if (index_lid < totalSize)
         estimationCalc(...);
   }
   // calculate the syndrome
   __syncthreads();
   for (p = 0; p < totalEdges; p += lgsize)
   {
      index_lid = p + lid;
      if (index_lid < totalSize)
      syndromeCalc(edgesFromCheck, estimation, syndromeOut, glPageStartIndex, index_e);
   }
   __syncthreads();   
   int cnodes = codeInfo[0].checkNodes;
   int parity = isAllZero(syndromeOut, cnodes, glPageStartIndex);
   // if syndrome is ok, decoding is successful
   if (parity == 1) return;
   __syncthreads();   
   int i;
   // forward and back iterations
   for (i = 0; i < iterations; i++)
   {
   // iteration to check nodes
      for (p = 0; p < totalEdges; p += lgsize)
      {
         index_lid = p + lid;
         if (index_lid < totalSize)
            iterateToCheck(...);
      }
      __syncthreads();
      . . .
      // iteration back to variable nodes
      __syncthreads();
      . . .
      // calculate the estimation
      __syncthreads();
      . . .
      // calculate the syndrome
      __syncthreads();
      // if syndrome is ok, decoding is successful
      parity = isAllZero(syndromeOut, cnodes, glPageStartIndex);
      if (parity == 1) return;
   }
}
\end{lstlisting}
\end{sublisting}}}}

\centering
\fcolorbox{black}{background}{\usebox\MyFirstListing}
\end{listing}

\end{document}